\documentclass[aps,prb,twocolumn,superscriptaddress]{revtex4-2}
\usepackage{graphicx}
\usepackage{bm}
\usepackage{amsmath,amssymb,amsfonts,bm}
\usepackage{epsfig}
\usepackage{epstopdf}
\usepackage{dcolumn}
\usepackage{grffile}
\usepackage{verbatim}
\usepackage{mathrsfs}
\usepackage{amssymb}
\usepackage{wasysym}
\usepackage{color,xcolor}
\usepackage{ulem}
\usepackage{extarrows}

\begin{document}
	
\title{Phenomenological Ginzburg-Landau theory for triple-Q magnetic orders on a hexagonal lattice}
	
\author{Jin-Tao Jin}
\affiliation{Department of Physics, The Hong Kong University of Science and Technology, Clear Water Bay, Kowloon 999077, Hong Kong, China}
	
\author{Yi Zhou}
\email{yizhou@iphy.ac.cn}
\affiliation{Institute of Physics, Chinese Academy of Sciences, Beijing 100190, China}
	
\begin{abstract}
We develop a comprehensive Ginzburg-Landau theory describing triple-Q magnetic orders on hexagonal lattices, focusing on $O(N)$ models with $N=2$ and $N=3$. Through systematic analysis of symmetry-allowed terms in the free energy, we establish complete phase diagrams governed by competing interaction parameters. Our theory reveals distinct magnetic configurations including single-Q, double-Q, and triple-Q states, each characterized by unique symmetry breaking patterns and collective excitations. The framework provides fundamental insights into complex magnetic orders recently observed in materials such as Na$_2$Co$_2$TeO$_6$, where the interplay between geometric frustration and multiple ordering vectors leads to exotic magnetic states. Our results establish clear connections between microscopic interactions, broken symmetries, and experimentally observable properties, offering a powerful tool for understanding and predicting novel magnetic phases in frustrated magnets.
\end{abstract}
	
\maketitle

\section{Introduction}\label{sec:introduction}
The Ginzburg-Landau (GL) theory serves as a cornerstone in understanding classical phase transitions and ordered states in condensed matter systems~\cite{ginzburg1950theory,landau1937theory}. Through systematic application of symmetry principles and order parameter expansions, the GL framework provides crucial insights into systems where competing interactions lead to complex phase diagrams and sophisticated ordering configurations.

In magnetic systems, the interplay between geometric frustration, spin-orbit coupling, and anisotropic exchange interactions gives rise to intricate multiple-Q magnetic structures that extend beyond conventional single-Q arrangements~\cite{Batista2016,HayamiPRB2021}. These multiple-Q configurations are particularly significant in hexagonal lattices, where they can generate novel topological features and unusual transport phenomena~\cite{Martin2010PRL,Kato2010PRL,Kumar2010}. Recent investigations of neutron scattering and thermal transport have revealed detailed magnetic patterns in materials such as Na$_2$Co$_2$TeO$_6$~\cite{YaoPRR2023,JinArxiv2024,Chen2023PRL,YaoPRL2022}, as also indicated in the Kitaev candidate material $\alpha$-RuCl$_3$~\cite{BanerjeeScience2017,SearsPRB2017,CaoPRB2016,Zhang2024PRM}. Similar complex ordering patterns have been observed in Na$_3$Co$_2$SbO$_6$~\cite{Gu2024PRB,LiPRX2022} and related compounds~\cite{Park2024Arxiv,Paddison2024NPJ,ParkNC2023}.

Of particular importance is the emergence of triple-Q states, which represent a distinct class of magnetic order stabilized by competition among different exchange interactions~\cite{KatoPRB2022}. While both theoretical studies~\cite{Messio2011,Janssen2016,Chen2023PRL,Francini2024PRB} and experimental observations~\cite{YaoPRR2023,JinArxiv2024} have highlighted the significance of these states, a comprehensive theoretical framework capable of explaining their formation and stability remains an outstanding challenge.

In this study, we develop a comprehensive GL theory for triple-Q magnetic orders on a hexagonal lattice. We define three order parameters, $\mathbf{\Delta}_{\bm{Q}_l}$ ($l=1,2,3$), each corresponding to a pair of M points within the Brillouin zone, with each $\mathbf{\Delta}_{\bm{Q}_l}$ representing an O($N$) multiplet. This approach provides sufficient generality to encompass a range of magnetic orders while remaining strictly governed by the inherent $D_3$ lattice symmetry~\cite{LiuPRB2018,SanoPRB2018,LiuPRL2020,Kim_2021}. Through detailed examination of system stability and various phases via GL free energy minimization, we focus particularly on the experimentally relevant cases of $N=2$ and $N=3$~\cite{Hayami2023PRB,YambePRB2023}. Our analysis yields comprehensive phase diagrams and clarifies the distinct configurations of order parameters under different conditions. Furthermore, we investigate collective excitations and symmetry breaking phenomena associated with different phases~\cite{BanerjeeNPJ2018,Samarakoon2021PRB,Hong2024NPJ,Miao2024PRB,Wang2023PRB,Pohle2023PRR}.

Our theoretical framework directly connects to recent advances in Kitaev candidate materials~\cite{TakagiNRP2019,TrebstPRRSP2022,MotomeJPCM2020} and frustrated magnetic systems~\cite{BalentsNature2010,ZhouRMP2017,BroholmScience2020}. The results provide clear directions for interpreting new experimental data and establish a foundation for potential applications in spintronics and quantum information technology~\cite{Winter2022,LiuIJMPB2021}.

\begin{figure*}[tb]
	\includegraphics[width=1.0\linewidth]{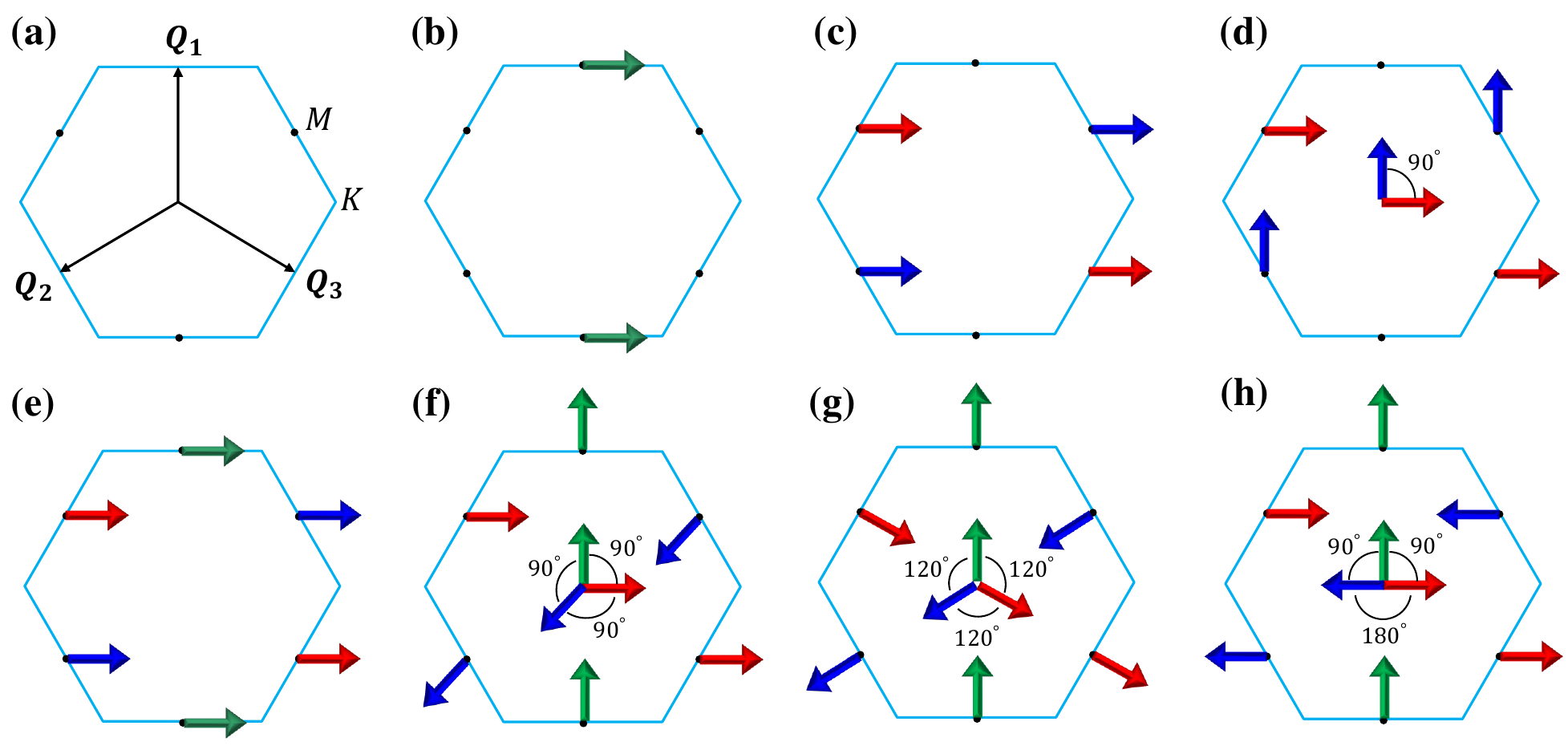}
	\caption{(a) The triple Q-vectors $\bm{Q}_1$, $\bm{Q}_2$, and $\bm{Q}_3$ are specified at the $M$ points of the first Brillouin Zone on a hexagonal lattice. Note that $-\bm{Q}_{l}$ is equivalent to $\bm{Q}_{l}$ at each $M$ point, where $l=1,2,3$. Multiple Q-vector magnetic configurations for $O(N)$ models include: single-Q (b) $\mathrm{I}_{O(N=2,3)}$, double-Q (c) $\mathrm{II}^{A}_{O(N=2,3)}$ and (d) $\mathrm{II}^{B}_{O(N=2,3)}$, and triple-Q (e) $\mathrm{III}^{A}_{O(N=2,3)}$ (collinear), (f) $\mathrm{III}^{B}_{O(3)}$ (orthogonal), (g) $\mathrm{III}^{C}_{O(2)}$ (120$^{\circ}$ coplanar), and (h) $\mathrm{III}^{D}_{O(2)}$ (orthogonal-collinear) states. The vectors that are collinear at distinct $\bm{Q}_l$ in (c), (e) and (h) may align in parallel or opposite directions. The configuration in (f) has three mutually orthogonal vectors, which is allowed only if $N\geq{}3$. The configurations in (g) and (h) are coplanar states.}\label{fig:QVec}
\end{figure*}

\section{Model and Symmetry}\label{sec:model- symmetry}
To construct an effective field theory for triple-Q magnetic orders on hexagonal lattices, we consider an $O(N)$ model with order parameters represented as $N$-dimensional vectors:
$$\bm{\Delta}_{\bm{Q}_l}=(\Delta^{1}_{\bm{Q}_l},\Delta^{2}_{\bm{Q}_l},\cdots,\Delta^{N}_{\bm{Q}_l})^{T},$$ 
where each component $\Delta^{p}_{\bm{Q}_l}$ is real. These order parameters correspond to the three $M$-points in the first Brillouin zone denoted by wave vectors: $\bm{Q}_1 = Q(0,1)$, $\bm{Q}_{2}=Q(-1/2,-1/2\sqrt{3})$, and $\bm{Q}_{3}=Q(1/2,-1/2\sqrt{3})$, as illustrated in Fig.~\ref{fig:QVec}(a).

The symmetry constraints of our system include the $D_3$ point group of the hexagonal lattice~\cite{Koster-Point-Groups} and the internal $O(N)$ rotational symmetry of the order parameters. Incorporating these symmetries, we express the Ginzburg-Landau free energy up to quartic order as~\cite{Sigrist-RMP}:
\begin{widetext}
\begin{equation}\label{eq:O-N-F-0}
\mathcal{F}_{O(N)}  =  \alpha \sum_{l=1}^{3} |\bm{\Delta}_{\bm{Q}_l}|^2 + \beta_1 \sum_{l=1}^{3} |\bm{\Delta}_{\bm{Q}_l}|^4 + \beta_2 \sum_{l<m}|\bm{\Delta}_{\bm{Q}_l}|^2|\bm{\Delta}_{\bm{Q}_m}|^2 + \beta_3 \sum_{l<m}\left(\bm{\Delta}_{\bm{Q}_l} \cdot \bm{\Delta}_{\bm{Q}_m}\right)^2,
\end{equation} 
\end{widetext}
where $\alpha$ controls the phase transition temperature and $\beta_{\lambda=1,2,3}$ are interaction parameters that determine the specific ordered phase.

For general parameter values, the free energy exhibits $D_3 \times O(N) \times \mathbb{Z}_2 \times \mathbb{Z}_2$ symmetry. The $D_3$ symmetry corresponds to the lattice point group, while $O(N)$ represents internal rotational symmetry. The additional $\mathbb{Z}_2$ factors arise from the reflection symmetry of each ordering vector.
This symmetry can be enhanced under specific parameter conditions: (1) When $\beta_3 = 0$, the system exhibits $D_3 \times O(N) \times O(N) \times O(N)$ symmetry, allowing independent rotations of each $\bm{\Delta}_{\bm{Q}_l}$. (2) When $\beta_2=2\beta_1$ and $\beta_3=0$, the symmetry reaches its maximum $O(3N)$, treating all components of all order parameters equivalently. 
Indeed, these enlarged symmetries suggest a phase transition between two distinct phases.

For systems with $N > 3$, we can always perform an $O(N)$ rotation to restrict nonzero components to at most three dimensions, making the $N=3$ analysis sufficient for understanding all cases with $N > 3$.
Hereafter we will focus on the experimentally relevant cases of $N=2$ (XY spins) and $N=3$ (Heisenberg spins), which capture the essential physics of most magnetic materials on hexagonal lattices. For example, Kitaev candidate materials featuring significant trigonal distortion along with direct exchange interactions can be modeled using an easy-plane XXZ model~\cite{Das2021,Maksimov2022,Winter2022}, which is effectively represented by the $O(2)$ model in low energies.

Specifically, in the spin density wave (SDW) content, we can explicitly define $$\bm{\Delta}_{\bm{Q}_l}\equiv\frac{1}{\sqrt{V}} \sum_{\bm{r}}\bm{S}(\bm{r}) e^{-i\bm{Q}_l\cdot{r}},$$ where $\bm{r}$ runs over the lattice sites, $V$ is the volume of the system, and $\bm{S}(\bm{r})$ is the local spin density. This definition is formally the same as the SDW order parameters introduced in Ref.~\cite{Nagaosa_CDW_SDW}. In addition, such magnetic order parameters can be explicitly defined in itinerant electron systems
 as 
\begin{equation*}
\begin{aligned}
\bm{\Delta}_{\bm{Q}_l}=\langle\bm{S}(\bm{Q}_l)\rangle
&=\frac{1}{2}\sum_{\bm{k}}(\langle{}c^\dagger_{\bm{k}\uparrow}c_{\bm{k}+\bm{Q}_l\downarrow}+c^\dagger_{\bm{k}\downarrow}c_{\bm{k}+\bm{Q}_l\uparrow}\rangle,\\
&\qquad\quad{-i}\langle{}c^\dagger_{\bm{k}\uparrow}c_{\bm{k}+\bm{Q}_l\downarrow}-c^\dagger_{\bm{k}\downarrow}c_{\bm{k}+\bm{Q}_l\uparrow}\rangle,\\
&\qquad\qquad\langle{}c^\dagger_{\bm{k}\uparrow}c_{\bm{k}+\bm{Q}_l\uparrow}-c^\dagger_{\bm{k}\downarrow}c_{\bm{k}+\bm{Q}_l\downarrow}\rangle)^{T},
\end{aligned}
\end{equation*}
where $\langle\dots\rangle$ denotes the expectation value. Our theoretical framework also accommodates potential competition between local moment and itinerant electron descriptions, which may emerge in certain materials—provided they adhere to the same symmetry configurations. However, the primary objective of our work is to uncover the universal physical properties governed by common symmetries, rather than to simulate material-specific details. As a result, while the field $\bm{\Delta_{\bm{Q}_l}}$ may describe charge density wave [$O(2)\cong{}U(1)$], SDW (N=3), magnetic insulator where local magnetic moments dominate over itinerant electrons, or more general O(N) orders, all such cases share the same symmetry-allowed quartic invariants and hence the same universal phase structure.

\begin{figure}[tb]
\includegraphics[width=\linewidth]{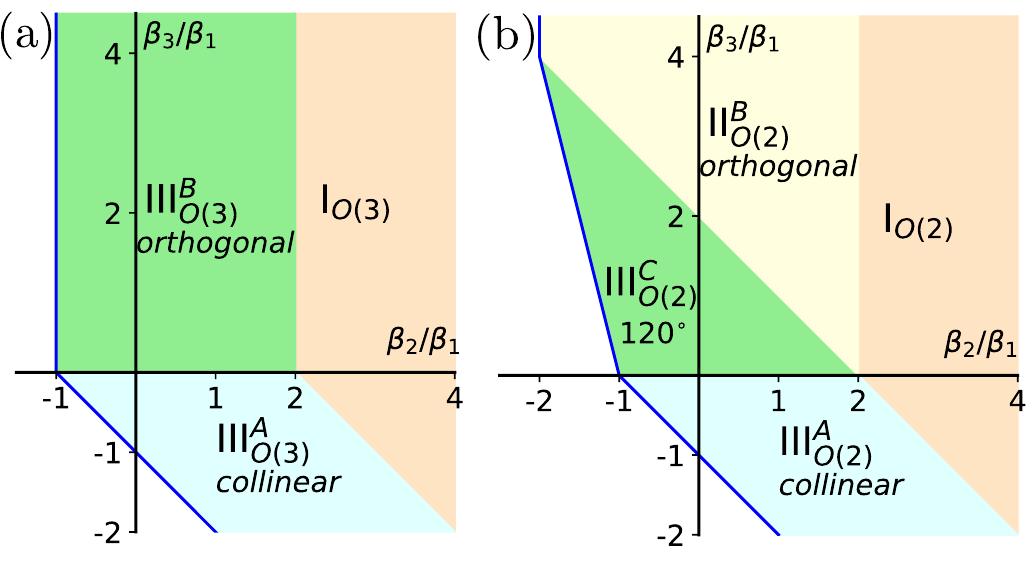}
\caption{
Phase diagrams for (a) $O(3)$ and (b) $O(2)$ models. Notations I, II, and III denote single-, double-, and triple-Q phases, respectively. Labels $A$, $B$, and $C$ indicate collinear, orthogonal, and 120$^\circ$ phases. Model (a) $O(3)$ is stable under conditions $\beta_1>0$, $\beta_1+\beta_2>0$, and $\beta_1+\beta_2+\beta_3>0$, while model (b) $O(2)$ is stable with $\beta_1>0$, $2\beta_1+\beta_2>0$, $\beta_1+\beta_2+\beta_3>0$, and $4(\beta_1+\beta_2)+\beta_3>0$. Each phase has a unique magnetic configuration, displayed in Fig.~\ref{fig:QVec}(b)-(g).
}\label{fig:PhaseDiagram}
\end{figure}

\section{Phase Diagrams}\label{sec:phase diagrams}
By minimizing the free energy in Eq.~(\ref{eq:O-N-F-0}), we can identify the stable ordered phases and construct phase diagrams for the $O(N)$ model. The stability conditions require $\alpha < 0$ and $\beta_1 > 0$, with additional constraints on $\beta_2$ and $\beta_3$ specific to each phase.

While focusing on nontrivial solutions with at least one nonzero element $|\bm{\Delta}_{\bm{Q}_l}|$, these solutions can be classified into three categories based on the number of nonzero wave vectors $\bm{Q}_l$:

(1) \textbf{Type $\mathrm{I}$} (Single-Q): Only one of the three values $|\bm{\Delta}_{\bm{Q}_l}|$ is nonzero. Without loss of generality, we choose $|\bm{\Delta}_{\bm{Q}_1}|\neq 0$. The corresponding free energy $\mathcal{F}^{\mathrm{I}}_{O(N)}$ simplifies to:
\begin{equation}\label{eq-sm:F-A}
\mathcal{F}^{\mathrm{I}}_{O(N)} = \alpha |{\bm{\Delta}}_{{\bm{Q}}_1}|^2 + \beta_1 |{\bm{\Delta}}_{{\bm{Q}}_1}|^4.
\end{equation}
	
(2) \textbf{Type $\mathrm{II}$} (double-Q): Two of the three components $|\bm{\Delta}_{\bm{Q}_l}|$ are nonzero. Assuming $|\bm{\Delta}_{\bm{Q}_3}| = 0$, the free energy $\mathcal{F}^{\mathrm{II}}_{O(N)}$ takes the form:
\begin{equation}\label{eq-sm:F-B}
\begin{aligned}
\mathcal{F}^{\mathrm{II}}_{O(N)}&= \alpha \left(|\bm{\Delta}_{\bm{Q}_1}|^2+|\bm{\Delta}_{\bm{Q}_2}|^2\right)+\beta_1 \left(|\bm{\Delta}_{\bm{Q}_1}|^4
+|\bm{\Delta}_{\bm{Q}_2}|^4\right)\\
&+\beta_2|\bm{\Delta}_{\bm{Q}_1}|^2|\bm{\Delta}_{\bm{Q}_2}|^2 +\beta_3 \left(\bm{\Delta}_{\bm{Q}_1} \cdot \bm{\Delta}_{\bm{Q}_2}\right)^2.    
\end{aligned}
\end{equation}
	
(3) \textbf{Type $\mathrm{III}$} (triple-Q): All three components $|\bm{\Delta}_{\bm{Q}_l}|$ are nonzero. The free energy $\mathcal{F}^{\mathrm{III}}_{O(N)}$ maintains the full form as given in Eq.~\eqref{eq:O-N-F-0}.

\subsection{$O(3)$ model}

For type I state denoted as I$_{O(3)}$, the system exhibits a single vector $\bm{\Delta}_{\bm{Q}_1}$ with magnitude $|\bm{\Delta}_{\bm{Q}_1}| = \sqrt{-\alpha/2\beta_1}$ and arbitrary orientation. The corresponding free energy is $\mathcal{F}^{\mathrm{I}}_{O(3)}= -\alpha^2/4\beta_1$. 

For type II states, we observe that 
$$0 \leq \left(\bm{\Delta}_{\bm{Q}_1} \cdot \bm{\Delta}_{\bm{Q}_2}\right)^2 \leq |\bm{\Delta}_{\bm{Q}_1}|^2|\bm{\Delta}_{\bm{Q}_2}|^2,$$ 
where the equality holds when $\bm{\Delta}_{\bm{Q}_1}$ and $\bm{\Delta}_{\bm{Q}_2}$ are either orthogonal or collinear. The free energy $\mathcal{F}^{\mathrm{II}}_{O(3)}$ reaches its minimum under different conditions depending on the sign of $\beta_3$:

\begin{equation}\label{eq:typeII-orientation}
\begin{split}
\mbox{for }\beta_3<0:&\quad \bm{\Delta}_{\bm{Q}_1} \parallel \bm{\Delta}_{\bm{Q}_2};\\
\mbox{for }\beta_3>0:&\quad \bm{\Delta}_{\bm{Q}_1} \cdot \bm{\Delta}_{\bm{Q}_2}=0.   
\end{split}
\end{equation}
By solving
\begin{equation*}
\frac{\partial \mathcal{F}^{\mathrm{II}}_{O(3)}} {\partial |\bm{\Delta}_{\bm{Q}_1}|} = \frac{\partial \mathcal{F}^\mathrm{II}_{O(3)}} {\partial |\bm{\Delta}_{\bm{Q}_2}|}=0,    
\end{equation*}
we find
\begin{equation}
|\bm{\Delta}_{\bm{Q}_1}|=|\bm{\Delta}_{\bm{Q}_2}|=\sqrt{\frac{-\alpha}{2\beta_1+\beta_2+\Theta(-\beta_3)\beta_3}},
\end{equation}
where $\Theta(x)$ is the Heaviside step function. The corresponding free energy is
\begin{equation}
\mathcal{F}^{\mathrm{II}}_{O(3)}=-\frac{\alpha^2}{2\beta_1+\beta_2+\Theta(-\beta_3)\beta_3}.
\end{equation}
We denote the type II state with collinear vectors as II$^A_{O(3)}$ and the state with orthogonal vectors as II$^B_{O(3)}$.

For type III states, we apply similar analysis using the inequalities
\begin{equation*}
0 \leq \sum_{l<m}\left(\bm{\Delta}_{\bm{Q}_l} \cdot \bm{\Delta}_{\bm{Q}_m}\right)^2 \leq \sum_{l<m}|\bm{\Delta}_{\bm{Q}_l}|^2|\bm{\Delta}_{\bm{Q}_m}|^2.    
\end{equation*}
When $\beta_3<0$, the minimum free energy corresponds to three collinear vectors (phase III$^A_{O(3)}$), while for $\beta_3>0$, it corresponds to three mutually orthogonal vectors (phase III$^B_{O(3)}$). In both cases:
\begin{equation}
|\bm{\Delta}_{\bm{Q}_1}|=|\bm{\Delta}_{\bm{Q}_2}|=|\bm{\Delta}_{\bm{Q}_3}|=\sqrt{\frac{-\alpha}{2[{\beta_1+\beta_2+\Theta(-\beta_3)\beta_3}]}},
\end{equation}
with free energy
\begin{equation}
\mathcal{F}^{\mathrm{III}}_{O(3)}=-\frac{3\alpha^2}{4[\beta_1+\beta_2+\Theta(-\beta_3)\beta_3]}.
\end{equation}

The global minimum of $\mathcal{F}_{O(3)}$ must be one of these five local minima ($\mathrm{I}_{O(3)}$, $\mathrm{II}^{A,B}_{O(3)}$, or $\mathrm{III}^{A,B}_{O(3)}$). The phase diagram is then obtained from the comparison of free energy among them. 

For the $O(3)$ model relevant to Heisenberg spins, we find three distinct stable phases, as shown in Fig.~\ref{fig:PhaseDiagram}(a):\\

(1) \textbf{Phase $\mathbf{I}_{O(3)}$} (Single-Q): Only one ordering vector is active, i.e., $|\bm{\Delta}_{\bm{Q}_1}| \neq 0$ and $|\bm{\Delta}_{\bm{Q}_2}| = |\bm{\Delta}_{\bm{Q}_3}| = 0$.
	
(2) \textbf{Phase $\mathbf{III}^A_{O(3)}$} (Triple-Q Collinear): All three ordering vectors $\bm{\Delta}_{\bm{Q}_{l=1,2,3}}$ have equal amplitudes and parallel directions.
	
(3) \textbf{Phase $\mathbf{III}^B_{O(3)}$} (Triple-Q Orthogonal): All three ordering vectors have equal amplitudes but are mutually orthogonal.\\
Fig.~\ref{fig:QVec}(b, e, f) illustrate the vector configurations in each phase. Notably, neither $\mathrm{II}^A_{O(3)}$ nor $\mathrm{II}^B_{O(3)}$ appears as the unique global minimum for any parameter values. System stability requires $\beta_1>0$, $\beta_1+\beta_2>0$, and $\beta_1+\beta_2+\beta_3>0$.

\subsection{$O(2)$ model}
For $N=2$, we can directly obtain the type I and type II solutions from their $N=3$ counterparts. The solutions $\mathrm{I}_{O(2)}$, $\mathrm{II}^A_{O(2)}$, and $\mathrm{II}^B_{O(2)}$ correspond directly to $\mathrm{I}_{O(3)}$, $\mathrm{II}^A_{O(3)}$, and $\mathrm{II}^B_{O(3)}$. However, type III solutions require special consideration since three mutually orthogonal vectors cannot exist in two-dimensional space.

To find type III solutions for $N=2$, we express $\bm{\Delta}_{\bm{Q}_l}$ in polar form:
$$\bm{\Delta}_{\bm{Q}_l}=|\bm{\Delta}_{\bm{Q}_l}|(\cos\phi_l,\sin\phi_l)^{T},$$
yielding inner products
$$\bm{\Delta}_{\bm{Q}_l}\cdot\bm{\Delta}_{\bm{Q}_m}=|\bm{\Delta}_{\bm{Q}_l}|\cdot|\bm{\Delta}_{\bm{Q}_m}|\cos(\phi_l-\phi_m).$$
The free energy in this representation becomes:
\begin{equation}
\begin{aligned}
\mathcal{F}^{\mathrm{III}}_{O(2)} = & \alpha \sum_{l=1}^{3} |{\bm{\Delta}}_{{\bm{Q}}_l}|^2
+ \beta_1 \sum_{l=1}^{3} |{\bm{\Delta}}_{{\bm{Q}}_l}|^4\\
&+\sum_{l<m} \left(\beta_2+\beta_3\cos^2\Phi_{lm}\right) |{\bm{\Delta}}_{{\bm{Q}}_l}|^2|{\bm{\Delta}}_{{\bm{Q}}_m}|^2,
\end{aligned}
\end{equation}
where $\Phi_{lm}=\phi_l-\phi_m$ represents the relative angle between vectors.

Minimizing the free energy leads to two sets of GL equations:
\begin{equation}
\begin{aligned}
\alpha+2\beta_1|\bm{\Delta}_{\bm{Q}_l}|^2+\sum_{m\neq{}l} \left(\beta_2+\beta_3\cos^2\Phi_{lm}\right) |{\bm{\Delta}}_{{\bm{Q}}_m}|^2 & =0, \\
\sum_{m\neq{}l} |\bm{\Delta}_{\bm{Q}_m}|^2\sin2\Phi_{lm} & = 0,
\end{aligned}    
\end{equation}
where $l=1,2,3$. While these equations yield six conditions, only five are independent. Eliminating $|\bm{\Delta}_{\bm{Q}_l}|$ results in:
\begin{equation}
\sin2\Phi_{12}=\sin2\Phi_{23}=\sin2\Phi_{31}.   
\end{equation}
This constraint leads to three distinct classes of solutions:

(1) Collinear configuration: When $\cos 2\Phi_{12}=\cos 2\Phi_{23}=\cos 2\Phi_{31}=1$, all three vectors $\bm{\Delta}_{\bm{Q}_l}$ are collinear. This solution, analogous to $\mathrm{III}_{O(3)}^A$, is designated as $\mathrm{III}_{O(2)}^A$.
	
(2) 120$^\circ$ configuration: When $\cos 2\Phi_{12}=\cos 2\Phi_{23}=\cos 2\Phi_{31}=-1/2$, the three vectors form a symmetric arrangement with mutual angles of 120$^\circ$. This solution yields:
\begin{equation}
\begin{aligned}
|{\bm{\Delta}}_{{\bm{Q}}_1}|=&|{\bm{\Delta}}_{{\bm{Q}}_2}|=|{\bm{\Delta}}_{{\bm{Q}}_3}|=\sqrt{-\frac{2\alpha}{4(\beta_1+\beta_2)+\beta_{3}}},\\
\mathcal{F}^{\mathrm{III}}_{O(2)}=&-\frac{3\alpha^2}{4(\beta_1+\beta_2)+\beta_{3}}.
\end{aligned}
\end{equation}
We denote this coplanar solution as $\mathrm{III}^{C}_{O(2)}$, illustrated in Fig.~\ref{fig:QVec}(g).
	
(3) Orthogonal-collinear configuration: When two values of $\cos 2\Phi_{lm}$ equal $-1$ and one equals $1$, two vectors align while remaining orthogonal to the third. This configuration leads to:
\begin{equation}
\begin{aligned}
|{\bm{\Delta}}_{{\bm{Q}}_2}|=&|{\bm{\Delta}}_{{\bm{Q}}_3}|=\sqrt{-\frac{\alpha(2\beta_1-\beta_2)}{4\beta_1^2-2\beta_2^2+2\beta_1(\beta_2+\beta_3)}},\\[1mm]
|{\bm{\Delta}}_{{\bm{Q}}_1}|=&\sqrt{-\frac{\alpha(2\beta_1-\beta_2+\beta_3)}{4\beta_1^2-2\beta_2^2+2\beta_1(\beta_2+\beta_3)}},\\[1mm]
\mathcal{F}^{\mathrm{III}}_{O(2)}=&-\frac{\alpha^2(6\beta_1-3\beta_2+\beta_3)}{8\beta_1^2-4\beta_2^2+4\beta_1(\beta_2+\beta_3)}.
\end{aligned}
\end{equation}
This solution, denoted as $\mathrm{III}^{D}_{O(2)}$, appears in Fig.~\ref{fig:QVec}(h).

By comparing the free energies of all six candidate phases ($\mathrm{I}_{O(2)}$, $\mathrm{II}^{A,B}_{O(2)}$, and $\mathrm{III}^{A,C,D}_{O(2)}$), we can obtain the corresponding phase diagram. 

The $O(2)$ model, applicable to XY spins, exhibits a richer phase diagram with four stable phases, as illustrated in Fig.~\ref{fig:PhaseDiagram}(b):\\

(1) \textbf{Phase $\mathbf{I}_{O(2)}$} (Single-Q): Similar to the $O(3)$ case, only one ordering vector is present. 
	
(2) \textbf{Phase $\mathbf{II}^B_{O(2)}$} (Double-Q Orthogonal): Two ordering vectors with equal amplitudes and orthogonal directions.
	
(3) \textbf{Phase $\mathbf{III}^A_{O(2)}$} (Triple-Q Collinear): All three ordering vectors have equal amplitudes and parallel directions. 
	
(4) \textbf{Phase $\mathbf{III}^C_{O(2)}$} (Triple-Q 120$^\circ$ Coplanar): All three ordering vectors have equal amplitudes with 120$^\circ$ angles between them in the $O(2)$ plane.\\ 

Fig.~\ref{fig:QVec}(b, d, e, g) illustrate the vector configurations in each phase. Notably, neither $\mathrm{II}^A_{O(2)}$ nor $\mathrm{III}^D_{O(2)}$ emerges as the unique global minimum for any parameter values. System stability requires three conditions: $2\beta_1+\beta_2>0$, $\beta_1+\beta_2+\beta_3>0$, and $ 4(\beta_1+\beta_2)+\beta_3>0$.

\subsection{Real-space Configurations}

The identified phases correspond to distinct real-space magnetic configurations. The order parameter $\bm{\Delta}(\bm{r})$ in real space can be reconstructed from the three ordering vectors $\bm{\Delta}_{\bm{Q}_l}$ through the relation
$$\bm{\Delta}(\bm{r})=\sum_{l}\bm{\Delta}_{\bm{Q}_l}e^{i\bm{Q}_l\cdot\bm{r}},$$
where $\bm{r}$ denotes the position of a unit cell in the original lattice. Using the primitive lattice vectors $\bm{a}_1=a(-1/2, \sqrt{3}/2)$ and $\bm{a}_2=a(-1, 0)$, we express the position as $\bm{r}=m\bm{a}_1+n\bm{a}_2$. Since the three M-point order parameters generate a 2×2 supercell structure in real space, the following four values of $\bm{\Delta}(\bm{r})$ within the new enlarged unit cell fully characterize the real-space magnetic order,
\begin{equation*}
\begin{aligned}
\bm{\Delta}(\bm{0})&= \bm{\Delta}_{\bm{Q}_1}+\bm{\Delta}_{\bm{Q}_2}+\bm{\Delta}_{\bm{Q}_3},\\
\bm{\Delta}(\bm{a}_1)&= -\bm{\Delta}_{\bm{Q}_1}+\bm{\Delta}_{\bm{Q}_2}-\bm{\Delta}_{\bm{Q}_3},\\
\bm{\Delta}(\bm{a}_2)&= \bm{\Delta}_{\bm{Q}_1}-\bm{\Delta}_{\bm{Q}_2}-\bm{\Delta}_{\bm{Q}_3},\\
\bm{\Delta}(\bm{a}_1+\bm{a}_2)&= -\bm{\Delta}_{\bm{Q}_1}-\bm{\Delta}_{\bm{Q}_2}+\bm{\Delta}_{\bm{Q}_3}.
\end{aligned}
\end{equation*}
The above expressions give rise to two key characteristics of $\bm{\Delta}(\bm{r})$: (i) The magnitude of $\bm{\Delta}(\bm{r})$ is not necessarily constant across different $\bm{r}$;
(ii) The vector sum of $\bm{\Delta}(\bm{r})$ over a $2\times 2$ supercell vanishes, i.e., $\bm{\Delta}(\bm{0})+\bm{\Delta}(\bm{a}_1)+\bm{\Delta}(\bm{a}_2)+\bm{\Delta}(\bm{a}_1+\bm{a}_2)=\bm{0}$. For simplicity, we consider the order parameter on a triangular lattice and examine the real-space configurations of $\bm{\Delta}(\bm{r})$ corresponding to each phase presented in Fig.~\ref{fig:PhaseDiagram}. 

\begin{figure*}[tb]
\includegraphics[width=0.8\linewidth]{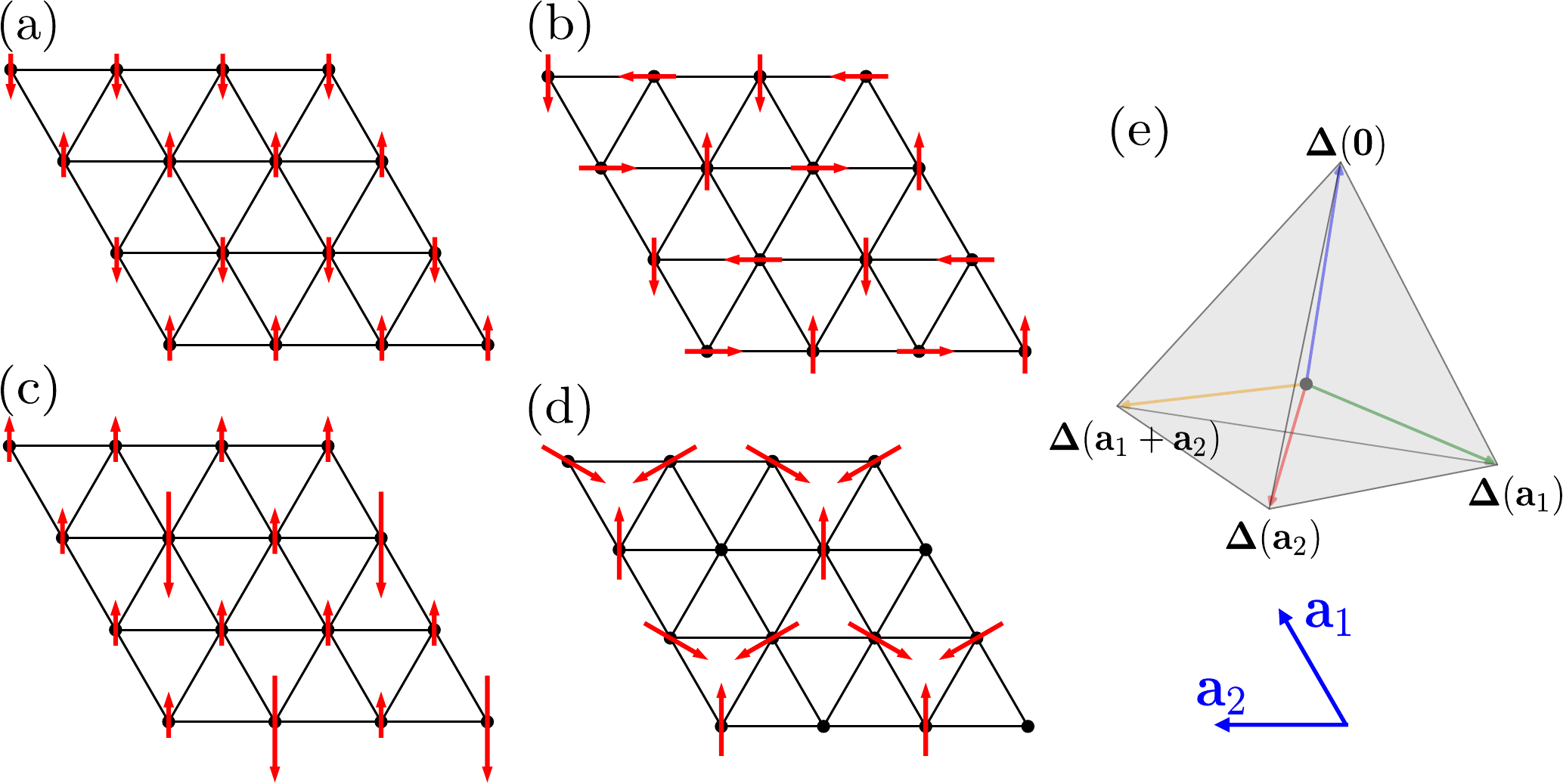}
	\caption{Real-space configurations of the order parameters for the phases (a) $\mathrm{I}_{O(N=2,3)}$, (b) $\mathrm{II}^B_{O(2)}$, (c) $\mathrm{III}^{A}_{O(N=2,3)}$, (d) $\mathrm{III}^{C}_{O(2)}$ and (e) $\mathrm{III}^B_{O(3)}$ in Fig.~\ref{fig:PhaseDiagram}. }\label{fig:real-space}
\end{figure*}

For the single-Q phase $\mathrm{I}_{O(N=2,3)}$, the magnetic structure forms a SDW along one direction and the supercell structure is reduced to $2\times{}1$, as shown in Fig.~\ref{fig:real-space}(a). This configuration is known as the ``row-wise antiferromagnetic state" \cite{Kurz2001} and has been studied in Ref.~\cite{Akagi2010} as well. The real-space configuration of the double-Q phase $\mathrm{II}^{B}_{O(2)}$
is presented in Fig.~\ref{fig:real-space}(b), which exhibits a coplanar orthogonal structure, also reported in Ref.~\cite{Akagi2010}. The real-space configuration of the triple-Q phase $\mathrm{III}^A_{O(N=2,3)}$ is shown in Fig.~\ref{fig:real-space}(c), corresponding to a 3:1 collinear arrangement. A similar structure appears in Ref.~\cite{Akagi2010}. However, a key difference is that in their case, the order parameter at every site has the same magnitude.

Particularly interesting is the triple-Q 120$^\circ$ phase in the $O(2)$ model, which forms a superposition of three SDWs with 120$^\circ$ angles between neighboring spins. Its configuration in Fig.~\ref{fig:real-space}(d) reveals a partially ordered state\cite{Ishitobi2023,Hattori2023,Hattori2024}, characterized by the absence of in-plane magnetic order at site $\bm{0}$. This configuration closely resembles the magnetic structure observed in materials such as Na$_2$Co$_2$TeO$_6$ and related compounds.

Furthermore, our theory identifies a non-coplanar spin structure for the triple-Q orthogonal phase $\mathrm{III}^B_{O(3)}$. The corresponding configuration displayed in Fig.~\ref{fig:real-space}(e) shows that the four $\bm{\Delta}(\bm{r})$ vectors within a supercell form a tetrahedral structure\cite{Akagi2010}, which can potentially give rise to non-trivial topological properties.

\section{Collective Excitations}\label{sec:collective excitations}
In the ordered phases, spontaneous symmetry breaking gives rise to low-energy collective modes. To analyze these excitations, we consider small fluctuations around the mean-field order parameter $\bm{\Delta}_{\bm{Q}_l}^0$:
\begin{equation}
\bm{\Delta}_{{\bm{Q}}_l}(\bm{r},t)=\bm{\Delta}^{0}_{{\bm{Q}}_l}+\bm{\eta}_{\bm{Q}_l}(\bm{r},t),
\end{equation}
where $|\bm{\eta}_{\bm{Q}_l}|\ll |\bm{\Delta}^{0}_{{\bm{Q}}_l}|$. Expanding the free energy to quadratic order in $\bm{\eta}_{\bm{Q}_l}$, we have
\begin{widetext}
\begin{equation}\label{eq:F-variation}
\begin{split}
\mathcal{F}_{O(N)} =  &\,\, \mathcal{F}^{0}_{O(N)} +\alpha\sum^{3}_{l=1}|\bm{\eta}_{\bm{Q}_l}|^2
+2\beta_1\sum^{3}_{l=1}\left[2(\bm{\Delta}^{0}_{\bm{Q}_l}\cdot\bm{\eta}_{\bm{Q}_l})^2+|\bm{\Delta}^{0}_{\bm{Q}_l}|^2|\bm{\eta}_{\bm{Q}_l}|^2 \right]\\
&+\beta_2\sum_{l<m}\left[4(\bm{\Delta}^{0}_{\bm{Q}_l}\cdot\bm{\eta}_{\bm{Q}_l})(\bm{\Delta}^{0}_{\bm{Q}_m}\cdot\bm{\eta}_{\bm{Q}_m})+|\bm{\Delta}^{0}_{\bm{Q}_l}|^2|\bm{\eta}_{\bm{Q}_m}|^2 +|\bm{\Delta}^{0}_{\bm{Q}_m}|^2|\bm{\eta}_{\bm{Q}_l}|^2\right]\\
&+\beta_3\sum_{l<m}\left[(\bm{\Delta}^{0}_{\bm{Q}_l}\cdot\bm{\eta}_{\bm{Q}_m})^2+(\bm{\Delta}^{0}_{\bm{Q}_m}\cdot\bm{\eta}_{\bm{Q}_l})^2+2(\bm{\Delta}^{0}_{\bm{Q}_l}\cdot\bm{\eta}_{\bm{Q}_m})(\bm{\Delta}^{0}_{\bm{Q}_m}\cdot\bm{\eta}_{\bm{Q}_l}) +2(\bm{\Delta}^{0}_{\bm{Q}_l}\cdot\bm{\Delta}^{0}_{\bm{Q}_m})(\bm{\eta}_{\bm{Q}_m}\cdot\bm{\eta}_{\bm{Q}_l})\right],
\end{split}
\end{equation} 
where $\mathcal{F}^{0}_{O(N)}$ is the free energy of the equilibrium state. 
To describe the temporal and spatial evolution of these fluctuations, we construct an effective Lagrangian that preserves all the symmetries inherent to the system, namely: (i) lattice translational symmetry, (ii) internal $O(N)\times\mathbb{Z}_2\times\mathbb{Z}_2$ symmetry, and (iii) $D_3$ lattice symmetry:
\begin{equation}\label{eq:effective-lagrangian}
\begin{split}
\mathcal{L}_{O(N)}&=
\Gamma\sum_{l}(\partial_t\bm{\eta}_{\bm{Q}_l})\cdot(\partial_t\bm{\eta}_{\bm{Q}_l}) - \kappa\sum_{l}\left[(\partial_x\bm{\eta}_{\bm{Q}_l})\cdot(\partial_x\bm{\eta}_{\bm{Q}_l})+(\partial_y\bm{\eta}_{\bm{Q}_l})\cdot(\partial_y\bm{\eta}_{\bm{Q}_l})\right]-\mathcal{F}_{O(N)},
\end{split}
\end{equation}
\end{widetext}
where $\Gamma$ and $\kappa$ are positive constants that represent the kinetic and spatial coupling parameters, respectively. The dynamics of the fluctuations are governed by the Euler-Lagrange equation: 
\begin{equation}\label{eq:collective-mode-equation}
-2\Gamma\partial_t^{2}\bm{\eta}_{\bm{Q}_l}+2\kappa(\partial^2_x+\partial^2_y)\bm{\eta}_{\bm{Q}_l}=\frac{\partial[\delta\mathcal{F}_{O(N)}]}{{\partial\bm{\eta}_{\bm{Q}_l}}}.
\end{equation}
Appendix \ref{app:effective Lagrangian} contains the derivation of both the symme-
try allowed Lagrangian and the Euler-Lagrange equation.

In the remainder of this section, we analyze the collective excitations for different phases based on Eq.~\eqref{eq:collective-mode-equation}. For convenience, we introduce the differential operator $$\square\equiv -2\Gamma\partial_t^{2}+2\kappa(\partial^2_x+\partial^2_y)$$ and set $|\bm{\Delta}^{0}_{\bm{Q}_l}|\equiv{}\Delta$ for all nonzero order parameters. For plane wave solutions of the form $\tilde{\eta}_i \propto \cos(\bm{q}\cdot\bm{r}-\omega{}t)$, the energy dispersion relation becomes:
\begin{equation}\label{eq:omega}
\omega^2=\frac{2\kappa{}q^2+\gamma_i(\alpha,\beta_1,\beta_2,\beta_3)}{2\Gamma},    
\end{equation}
where $\gamma_i$ represents a non-negative function of coefficients $\alpha$ and $\beta_{\lambda=1,2,3}$. We want to obtain $\gamma_i$ in Eq.~\eqref{eq:omega} for each $\tilde{\eta}_i$. This gives rise to the energy gap $\sqrt{\gamma_i/2\Gamma}$ in the long-wavelength limit ($q\rightarrow{}0$).

\subsection{Collective excitations for $O(2)$ model}

We explore the collective excitations for the $N=2$ scenario by studying each phase sequentially from the simplest to the most complex. For every instance, the fluctuations around the equilibrium state result in both gapped and gapless modes, whose characteristics are closely associated with symmetry breaking.

\textbf{Phase $\mathrm{I}_{O(2)}$:}
In the simplest case, the equilibrium is characterized by a single order parameter. We set 
$$
\bm{\Delta}^{\,0}_{\bm{Q}_1}=\Delta\,(1,0)^T,
$$
where $\Delta=\sqrt{-\alpha/2\beta_1}$, and decompose the fluctuation as 
$$
\bm{\eta}_{\bm{Q}_1}=(\eta^1_{\bm{Q}_1},\,\eta^2_{\bm{Q}_1})^T.
$$
This leads to two decoupled differential equations:
\begin{subequations}\label{eq-sm:collective-A-N=2}
\begin{align}
\square{}\eta^1_{\bm{Q}_1}&=-4\alpha\eta^1_{\bm{Q}_1},\\
\square{}\eta^2_{\bm{Q}_1}&=0.
\end{align}
\end{subequations}
In this context, the modes 
$$
\tilde{\eta}_1=\eta^1_{\bm{Q}_1}\quad\text{and}\quad \tilde{\eta}_2=\eta^2_{\bm{Q}_1}
$$
demonstrate a gapped excitation with a gap of $\gamma_1=-4\alpha$ and a gapless excitation characterized by $\gamma_2=0$, respectively. The presence of the gapless mode indicates the occurrence of a Goldstone mode due to the breaking of the internal $O(2)$ symmetry.

\textbf{Phase $\mathrm{II}^B_{O(2)}$:}
Subsequently, we analyze a phase characterized by the presence of two orthogonal order parameters. We utilize $$
\bm{\Delta}^{\,0}_{\bm{Q}_1}=\Delta\,(1,0)^{T}\quad\mbox{and}\quad \bm{\Delta}^{\,0}_{\bm{Q}_2}=\Delta\,(0,1)^{T},
$$ with $\Delta=\sqrt{\frac{-\alpha}{2\beta_1+\beta_2}}$, to represent this scenario and describe the fluctuations as $$
\bm{\eta}_{\bm{Q}_l}=(\eta^{1}_{\bm{Q}_l},\,\eta^{2}_{\bm{Q}_l})^T,\quad (l=1,2).
$$
The resulting equations of motion read:
\begin{equation}
\square 
\begin{pmatrix}
\eta_{\bm{Q}_1}^1 \\
\eta_{\bm{Q}_1}^2 \\
\eta_{\bm{Q}_2}^1 \\
\eta_{\bm{Q}_2}^2
\end{pmatrix}
= 2\Delta^2 
\begin{pmatrix}
4\beta_1 & 0 & 0 & 2\beta_2 \\
0 & \beta_3 & \beta_3 & 0 \\
0 & \beta_3 & \beta_3 & 0 \\
2\beta_2 & 0 & 0 & 4\beta_1
\end{pmatrix}
\begin{pmatrix}
\eta_{\bm{Q}_1}^1 \\
\eta_{\bm{Q}_1}^2 \\
\eta_{\bm{Q}_2}^1 \\
\eta_{\bm{Q}_2}^2
\end{pmatrix}    
\end{equation}

Straightforward algebra reveals that the four collective modes can be described as 
$$
\tilde{\eta}_{1,2}=\eta^1_{\bm{Q}_1}\pm \eta^{2}_{\bm{Q}_2}\quad\mbox{and}\quad \tilde{\eta}_{3,4}=\eta^2_{\bm{Q}_1}\pm \eta^{1}_{\bm{Q}_2},
$$
with the following gaps: 
$$
\gamma_1=-4\alpha,\,\,\gamma_2=4(2\beta_1-\beta_2)\Delta^2,\,\, \gamma_3=4\beta_3\Delta^2,\,\, \gamma_4=0.
$$
Once more, the appearance of the gapless mode is due to the broken $O(2)$ symmetry.

\textbf{Phase $\mathrm{III}^A_{O(2)}$:}
For a phase with three parallel order parameters, we choose 
$$
\bm{\Delta}^{\,0}_{\bm{Q}_l}=\Delta\,(1,0)^T, \quad l=1,2,3,
$$
where $\Delta=\sqrt{\frac{-\alpha}{2[{\beta_1+\beta_2+\beta_3}]}}$, and write the fluctuations as 
$$
\bm{\eta}_{\bm{Q}_l}=(\eta^1_{\bm{Q}_l},\,\eta^2_{\bm{Q}_l})^T.
$$
The equations of motion then separate into two independent sets:
\begin{subequations}\label{eq-sm:collective-C-1-N=2}
\begin{align}
\square 
\left(
\begin{matrix}
\eta^1_{\bm{Q}_1}\\ \eta^1_{\bm{Q}_2}\\ \eta^1_{\bm{Q}_3}
\end{matrix}
\right)
&=4\Delta^2
\left(
\begin{matrix}
2\beta_1 &\beta_2+\beta_3 &\beta_2+\beta_3\\
\beta_2+\beta_3 &2\beta_1 &\beta_2+\beta_3\\
\beta_2+\beta_3 &\beta_2+\beta_3 &2\beta_1
\end{matrix}
\right)
\left(
\begin{matrix}
\eta^1_{\bm{Q}_1}\\ \eta^1_{\bm{Q}_2}\\ \eta^1_{\bm{Q}_3}
\end{matrix}
\right),\\
\square 
\left(
\begin{matrix}
\eta^2_{\bm{Q}_1}\\ \eta^2_{\bm{Q}_2}\\ \eta^2_{\bm{Q}_3}
\end{matrix}
\right)
&=2\beta_3\Delta^2
\left(
\begin{matrix}
-2 &1 &1\\
1 &-2 &1\\
1 &1 &-2
\end{matrix}
\right)
\left(
\begin{matrix}
\eta^1_{\bm{Q}_1}\\ \eta^1_{\bm{Q}_2}\\ \eta^1_{\bm{Q}_3}
\end{matrix}
\right).
\end{align}
\end{subequations}
The linear combination defined as 
$$
\tilde{\eta}_1=\eta^1_{\bm{Q}_1}+\eta^1_{\bm{Q}_2}+\eta^1_{\bm{Q}_3}
$$
is associated with a gapped mode characterized by a gap $\gamma_1=-4\alpha$. The other two linear combinations involving the $\eta^1_{\bm{Q}_l}$ result in degenerate gapped modes with 
$$
\gamma_2=\gamma_3=4(2\beta_1-\beta_2-\beta_3)\Delta^2.
$$
In parallel, a gapless mode is produced in the $\eta^2_{\bm{Q}_l}$ domain, expressed as:
$$
\tilde{\eta}_6=\eta^2_{\bm{Q}_1}+\eta^2_{\bm{Q}_2}+\eta^2_{\bm{Q}_3} \quad (\gamma_6=0),
$$
while the two other independent combinations result in degenerate gapped modes given by 
$$
\gamma_4=\gamma_5=-6\beta_3\Delta^2.
$$

\begin{figure}[tb]
\includegraphics[width=0.6\linewidth]{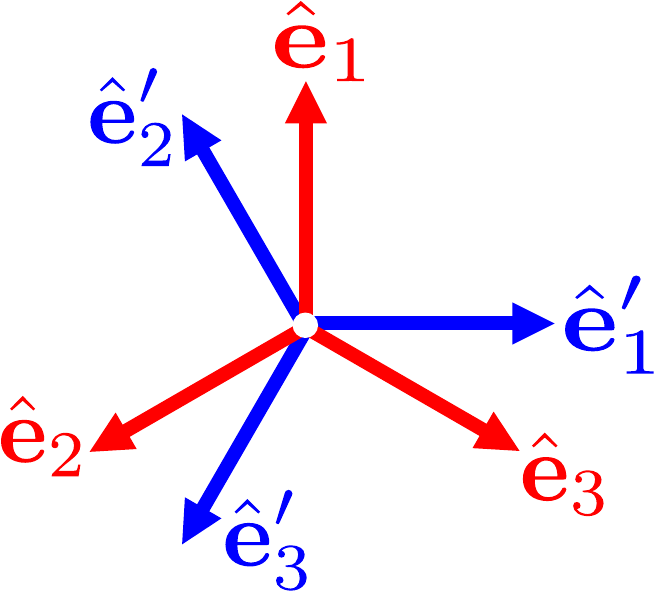}
\caption{Unit vectors that form the basis for the decomposition of $\bm{\Delta}_{\bm{Q}_l}$ and $\bm{\eta}_{\bm{Q}_l}$ for phase $\mathrm{III}^C_{O(2)}$.}\label{fig:unit-vectors-III-B-O(2)}
\end{figure}

\textbf{Phase $\mathrm{III}^C_{O(2)}$:}
In the phase characterized by the most intricate structure, where rotation of the order parameters occurs, each of the three parameters points in unique directions. Namely, we have
$$
\bm{\Delta}^{\,0}_{\bm{Q}_l}=\Delta\,\hat{\bm{e}}_l,
$$
where $\Delta=\sqrt{-\frac{2\alpha}{4(\beta_1+\beta_2)+\beta_{3}}}$, and the unit vectors are specified as follows:
$$
\hat{\bm{e}}_1=(0,1)^T,\,\, \hat{\bm{e}}_2=\Big(-\frac{\sqrt{3}}{2},-\frac{1}{2}\Big)^T,\,\, \hat{\bm{e}}_3=\Big(\frac{\sqrt{3}}{2},-\frac{1}{2}\Big)^T.
$$
It is useful to define perpendicular unit vectors in addition:
$$
\hat{\bm{e}}'_1=(1,0)^T,\,\, \hat{\bm{e}}'_2=\Big(-\frac{1}{2},\frac{\sqrt{3}}{2}\Big)^T,\,\, \hat{\bm{e}}'_3=\Big(-\frac{1}{2},-\frac{\sqrt{3}}{2}\Big)^T.
$$
These vectors, shown in Fig. 4, satisfy the orthogonality requirement $\hat{\bm{e}}_l\cdot\hat{\bm{e}}'_l=0$. For $l\neq m$, we also have
$$
\hat{\bm{e}}_l\cdot\hat{\bm{e}}_m=\hat{\bm{e}}'_l\cdot\hat{\bm{e}}'_m=-\frac{1}{2}\quad\mbox{and}\quad\hat{\bm{e}}_l\cdot\hat{\bm{e}}'_m=\frac{\sqrt{3}}{2}\,\epsilon_{lm},
$$
where $\epsilon_{lm}$ denotes the antisymmetric tensor. When the fluctuation field is expanded on this basis:
$$
\bm{\eta}_{\bm{Q}_l}=\eta^1_{\bm{Q}_l}\,\hat{\bm{e}}_l+\eta^2_{\bm{Q}_l}\,\hat{\bm{e}}'_l,
$$
it allows the derivative terms in the free energy to decouple along these specified directions.

Furthermore, the equilibrium and fluctuation fields yield the following relations:
\begin{align*}
\bm{\Delta}^{\,0}_{\bm{Q}_l}\cdot\bm{\Delta}^{\,0}_{\bm{Q}_m}&=-\frac{1}{2}\,\Delta^2 \quad (l\neq m),\\[1mm]
\bm{\Delta}^{\,0}_{\bm{Q}_l}\cdot\bm{\eta}_{\bm{Q}_l}&=\Delta\,\eta^1_{\bm{Q}_l},\\[1mm]
\bm{\Delta}^{\,0}_{\bm{Q}_l}\cdot\bm{\eta}_{\bm{Q}_m}+\bm{\Delta}^{\,0}_{\bm{Q}_m}\cdot\bm{\eta}_{\bm{Q}_l}&=-\frac{1}{2}\Delta\Big(\eta^1_{\bm{Q}_l}+\eta^1_{\bm{Q}_m}\Big)\\
&-\frac{\sqrt{3}}{2}\Delta\,\epsilon_{lm}\Big(\eta^2_{\bm{Q}_l}-\eta^2_{\bm{Q}_m}\Big).
\end{align*}

The equations of motion for the six fluctuation components can then be jointly written as
\begin{widetext}
\begin{equation}\label{eq-sm:collective-C-2-N=2}
\square
\begin{pmatrix}
\eta^1_{\bm{Q}_1} \\[1mm] \eta^1_{\bm{Q}_2} \\[1mm] \eta^1_{\bm{Q}_3} \\[1mm] \eta^2_{\bm{Q}_1} \\[1mm] \eta^2_{\bm{Q}_2} \\[1mm] \eta^2_{\bm{Q}_3}
\end{pmatrix}=
\Delta^2\,
\begin{pmatrix}
8\beta_1 & 4\beta_2+\beta_3 & 4\beta_2+\beta_3 & 0 & -\sqrt{3}\beta_3 & \sqrt{3}\beta_3 \\[1mm]
4\beta_2+\beta_3 & 8\beta_1 & 4\beta_2+\beta_3 & \sqrt{3}\beta_3 & 0 & -\sqrt{3}\beta_3 \\[1mm]
4\beta_2+\beta_3 & 4\beta_2+\beta_3 & 8\beta_1 & -\sqrt{3}\beta_3 & \sqrt{3}\beta_3 & 0 \\[1mm]
0 & \sqrt{3}\beta_3 & -\sqrt{3}\beta_3 & 2\beta_3 & -\beta_3 & -\beta_3\\[1mm]
-\sqrt{3}\beta_3 & 0 & \sqrt{3}\beta_3 & -\beta_3 & 2\beta_3 & -\beta_3\\[1mm]
\sqrt{3}\beta_3 & -\sqrt{3}\beta_3 & 0 & -\beta_3 & -\beta_3 & 2\beta_3
\end{pmatrix}
\begin{pmatrix}
\eta^1_{\bm{Q}_1} \\[1mm] \eta^1_{\bm{Q}_2} \\[1mm] \eta^1_{\bm{Q}_3} \\[1mm] \eta^2_{\bm{Q}_1} \\[1mm] \eta^2_{\bm{Q}_2} \\[1mm] \eta^2_{\bm{Q}_3}
\end{pmatrix}.
\end{equation}    
\end{widetext}
A mode analysis reveals that among the six collective modes, two can be directly identified:

(1) The gapped mode
$$
\tilde{\eta}_1=\eta^1_{\bm{Q}_1}+\eta^1_{\bm{Q}_2}+\eta^1_{\bm{Q}_3},\quad \gamma_1=-4\alpha,
$$
(2) The gapless (Goldstone) mode
$$
\tilde{\eta}_2=\eta^2_{\bm{Q}_1}+\eta^2_{\bm{Q}_2}+\eta^2_{\bm{Q}_3},\quad \gamma_2=0.
$$

The four remaining modes involve mixing between the $\eta^1_{\bm{Q}_l}$ and $\eta^2_{\bm{Q}_l}$ sectors and appear as two degenerate pairs. They can be written as:
\begin{widetext}
\begin{align*}
\tilde{\eta}_3 &= \frac{1}{\beta_1}\left[(\gamma_3/\Delta^2-3\beta_3)\Big(\eta^1_{\bm{Q}_1}+\eta^1_{\bm{Q}_2}-2\eta^1_{\bm{Q}_3}\Big)+3\sqrt{3}\,\beta_3\,\Big(\eta^2_{\bm{Q}_1}-\eta^2_{\bm{Q}_2}\Big)\right],\\[1mm]
\tilde{\eta}_4 &= \frac{1}{\beta_1}\left[(\gamma_4/\Delta^2-3\beta_3)\Big(\eta^1_{\bm{Q}_1}-2\eta^1_{\bm{Q}_2}+\eta^1_{\bm{Q}_3}\Big)-3\sqrt{3}\,\beta_3\,\Big(\eta^2_{\bm{Q}_1}-\eta^2_{\bm{Q}_3}\Big)\right],\\[1mm]
\tilde{\eta}_5 &= \frac{1}{\beta_1}\left[(\gamma_5/\Delta^2-3\beta_3)\Big(\eta^1_{\bm{Q}_1}+\eta^1_{\bm{Q}_2}-2\eta^1_{\bm{Q}_3}\Big)+3\sqrt{3}\,\beta_3\,\Big(\eta^2_{\bm{Q}_1}-\eta^2_{\bm{Q}_2}\Big)\right],\\[1mm]
\tilde{\eta}_6 &= \frac{1}{\beta_1}\left[(\gamma_6/\Delta^2-3\beta_3)\Big(\eta^1_{\bm{Q}_1}-2\eta^1_{\bm{Q}_2}+\eta^1_{\bm{Q}_3}\Big)-3\sqrt{3}\,\beta_3\,\Big(\eta^2_{\bm{Q}_1}-\eta^2_{\bm{Q}_3}\Big)\right],
\end{align*}    
\end{widetext}
with the eigenvalues given by
\begin{widetext}
\begin{subequations}\label{eq-sm:gap_IIIC}
\begin{eqnarray}
\gamma_3=\gamma_4 &= & \Delta^2\Big[2(2\beta_1-\beta_2)+\beta_3-\sqrt{4(2\beta_1-\beta_2)^2-8(2\beta_1-\beta_2)\beta_3+13\beta_3^2}\,\Big],\\[1mm]
\gamma_5=\gamma_6 &= & \Delta^2\Big[2(2\beta_1-\beta_2)+\beta_3+\sqrt{4(2\beta_1-\beta_2)^2-8(2\beta_1-\beta_2)\beta_3+13\beta_3^2}\,\Big].
\end{eqnarray}
\end{subequations}    
\end{widetext}

\subsection{Collective excitations for $O(3)$ model}

We proceed by extending our analysis to systems with $N=3$, using the understanding gained from the $N=2$ scenario as our basis. With a focus on the influence of the extra degree of freedom on the collective modes, we delve into each phase with growing complexity.

\textbf{Phase $\mathrm{I}_{O(3)}$:}
In the simplest phase involving only one order parameter, we define
$$
\bm{\Delta}^0_{\bm{Q}_1}=(1,0,0)^T
$$
where $\Delta=\sqrt{-\alpha/2\beta_1}$, and represent the fluctuation field by
$$
\bm{\eta}_{\bm{Q}_1}=(\eta^1_{\bm{Q}_1},\eta^2_{\bm{Q}_1},\eta^3_{\bm{Q}_1})^T.
$$
The equations of motion become fully decoupled:
\begin{subequations}\label{eq-sm:collective-A-N=3}
\begin{align}
\square{}\eta^1_{\bm{Q}_1}&=-4\alpha\eta^1_{\bm{Q}_1},\\[1mm]
\square{}\eta^2_{\bm{Q}_1}&=0,\\[1mm]
\square{}\eta^3_{\bm{Q}_1}&=0.
\end{align}
\end{subequations}
The collective modes correspond directly to the components: $\tilde{\eta}_i=\eta^{p=i}_{\bm{Q}_1} (i=1,2,3)$, where $\tilde{\eta}_1$ is gapped with $\gamma_1=-4\alpha$, while the modes $\tilde{\eta}_2$ and $\tilde{\eta}_3$ are gapless with $\gamma_2=\gamma_3=0$. The two gapless modes reflect the spontaneous breaking of the $O(3)$ symmetry to $O(2)$.

\textbf{Phase $\mathrm{III}^A_{O(3)}$:}
In the phase characterized by three parallel order parameters, we set
$$
\bm{\Delta}^0_{\bm{Q}_l}=\Delta(1,0,0)^T, \quad l=1,2,3,
$$
accompanied by fluctuations given by
$$
\bm{\eta}_{\bm{Q}_l}=(\eta^1_{\bm{Q}_l},\eta^2_{\bm{Q}_l},\eta^3_{\bm{Q}_l})^T,
$$ where $\Delta=\sqrt{\frac{-\alpha}{2[{\beta_1+\beta_2+\beta_3}]}}$. 
The expressions for $\eta^1_{\bm{Q}_l}$ and $\eta^2_{\bm{Q}_l}$ correspond directly to phase $\mathrm{III}^A_{O(2)}$, resulting in six collective modes ($\tilde{\eta}_1$ through $\tilde{\eta}_6$). Similarly, the expressions for $\eta^3_{\bm{Q}_l}$ are identical to those for $\eta^2_{\bm{Q}_l}$, which is attributable to the $O(2)$ symmetry present in the plane orthogonal to $\bm{\Delta}^0_{\bm{Q}_l}$. This yields three additional modes:

(1) Two degenerate gapped modes:
$$\tilde{\eta}_7=\eta^3_{\bm{Q}_1}-\eta^3_{\bm{Q}_2}, \quad \tilde{\eta}_8=\eta^3_{\bm{Q}_1}-\eta^3_{\bm{Q}_3}$$
with $\gamma_7=\gamma_8=-6\beta_3\Delta^2$;

(2) One gapless mode:
$$\tilde{\eta}_9=\eta^3_{\bm{Q}_1}+\eta^3_{\bm{Q}_2}+\eta^3_{\bm{Q}_3}$$
with $\gamma_9=0$.

In total, this phase exhibits seven gapped modes and two gapless modes.

\textbf{Phase III$^B_{O(3)}$:}
In the most complex configuration, the three order parameters are mutually perpendicular:
\begin{align*}
\bm{\Delta}^0_{\bm{Q}_1}=\Delta(1,0,0)^T,
\bm{\Delta}^0_{\bm{Q}_2}=\Delta(0,1,0)^T,
\bm{\Delta}^0_{\bm{Q}_3}=\Delta(0,0,1)^T,
\end{align*}
with fluctuations $\bm{\eta}_{\bm{Q}_l}=(\eta^1_{\bm{Q}_l},\eta^2_{\bm{Q}_l},\eta^3_{\bm{Q}_l})^T$, where $\Delta=\sqrt{\frac{-\alpha}{2(\beta_1+\beta_2)}}$. 

This configuration yields $\bm{\Delta}^0_{\bm{Q}_l}\cdot\bm{\eta}_{\bm{Q}_m}= \Delta\eta^l_{\bm{Q}_m}$. The nine components $\eta^p_{\bm{Q}_l}$ separate into four coupled sets:
\begin{subequations}\label{eq-sm:collective-C-2-N=3}
\begin{align}
\square 
\begin{pmatrix}
\eta^1_{\bm{Q}_1}\\ \eta^2_{\bm{Q}_2}\\ \eta^3_{\bm{Q}_3}
\end{pmatrix}
&=4\Delta^2
\begin{pmatrix}
2\beta_1 &\beta_2+\beta_3 &\beta_2\\
\beta_2 &2\beta_1 &\beta_2\\
\beta_2 &\beta_2 &2\beta_1
\end{pmatrix}
\begin{pmatrix}
\eta^1_{\bm{Q}_1}\\ \eta^2_{\bm{Q}_2}\\ \eta^3_{\bm{Q}_3}
\end{pmatrix},\\[3mm]
\square 
\begin{pmatrix}
\eta^2_{\bm{Q}_1}\\ \eta^1_{\bm{Q}_2}
\end{pmatrix}
&=2\beta_3\Delta^2
\begin{pmatrix}
1 &1\\
1 &1
\end{pmatrix}
\begin{pmatrix}
\eta^2_{\bm{Q}_1}\\ \eta^1_{\bm{Q}_2}
\end{pmatrix},\\[3mm]
\square 
\begin{pmatrix}
\eta^3_{\bm{Q}_1}\\ \eta^1_{\bm{Q}_3}
\end{pmatrix}
&=2\beta_3\Delta^2
\begin{pmatrix}
1 &1\\
1 &1
\end{pmatrix}
\begin{pmatrix}
\eta^3_{\bm{Q}_1}\\ \eta^1_{\bm{Q}_3}
\end{pmatrix},\\[3mm]
\square 
\begin{pmatrix}
\eta^3_{\bm{Q}_2}\\ \eta^2_{\bm{Q}_3}
\end{pmatrix}
&=2\beta_3\Delta^2
\begin{pmatrix}
1 &1\\
1 &1
\end{pmatrix}
\begin{pmatrix}
\eta^3_{\bm{Q}_2}\\ \eta^2_{\bm{Q}_3}
\end{pmatrix}.
\end{align}
\end{subequations}
Analysis of these coupled equations of motion leads to nine collective modes:

(1) One gapped mode with $\gamma_1=-4\alpha$:
$$\tilde{\eta}_1=\eta^1_{\bm{Q}_1}+\eta^2_{\bm{Q}_2}+\eta^3_{\bm{Q}_3}.$$

(2) Two degenerate gapped modes with $\gamma_2=\gamma_3=4(2\beta_1-\beta_2)\Delta^2$:
$$\tilde{\eta}_2=\eta^1_{\bm{Q}_1}-\eta^2_{\bm{Q}_2}, \quad \tilde{\eta}_3=\eta^1_{\bm{Q}_1}-\eta^3_{\bm{Q}_3}.$$

(3) Three degenerate gapped modes with $\gamma_4=\gamma_5=\gamma_6=4\beta_3\Delta^2$:
$$\tilde{\eta}_4=\eta^2_{\bm{Q}_1}+\eta^1_{\bm{Q}_2}, \quad \tilde{\eta}_5=\eta^3_{\bm{Q}_1}+\eta^1_{\bm{Q}_3}, \quad \tilde{\eta}_6=\eta^3_{\bm{Q}_2}+\eta^2_{\bm{Q}_3}.$$
	
(4) Three degenerate gapless modes with $\gamma_7=\gamma_8=\gamma_9=0$:
$$\tilde{\eta}_7=\eta^2_{\bm{Q}_1}-\eta^1_{\bm{Q}_2}, \quad	\tilde{\eta}_8=\eta^3_{\bm{Q}_1}-\eta^1_{\bm{Q}_3}, \quad \tilde{\eta}_9=\eta^3_{\bm{Q}_2}-\eta^2_{\bm{Q}_3}.$$

\subsection{Symmetry Breaking}
The symmetry breaking patterns from the initial $D_3 \times O(N) \times \mathbb{Z}_2 \times \mathbb{Z}_2$ group reveal distinct physical characteristics across different ordered phases.

For $N=3$: (1) $\mathrm{I}_{O(3)}$ phase breaks $D_3$ to $C_2$ and partially breaks $O(3)$, yielding residual symmetry $C_2 \times O(2)$. (2)  $\mathrm{III}^A_{O(3)}$ phase preserves $D_3$ while breaking internal symmetry to $O(2)$, resulting in $D_3 \times O(2)$. (3) In $\mathrm{III}^B_{O(3)}$  phase, both $D_3$ and internal symmetries break, but preserves octahedral group $O$ through combined operations.

For $N=2$: (1) $\mathrm{I}_{O(2)}$ phase breaks the initial symmetry to $C_2\times\mathbb{Z}_2$. (2) In $\mathrm{II}^B_{O(2)}$ phase, residual symmetry $C_4$ is generated by $C^{\prime}_{2}\times{}C_{4}\times{}(-1)\times{}1\,\, (\in{}D_3 \times O(2) \times \mathbb{Z}_2 \times \mathbb{Z}_2)$. (3) In $\mathrm{III}^A_{O(2)}$ phase, the initial symmetry breaks to $D_3\times{}\mathbb{Z}_2$. (4) For $\mathrm{III}^C_{O(2)}$ phase, combined lattice-internal operations preserve $D_3$.

The spontaneous breaking of continuous symmetries results in gapless Goldstone modes~\cite{Goldstone_Nambu,Goldstone_theorem,Goldstone_counting_Haruki}. The number of such modes equals the number of broken continuous symmetry generators. For example, in the single-Q phase of the $O(3)$ model, the reduction $O(3)\to O(2)$ produces two gapless modes with linear dispersion, $\omega(\bm{q})\sim \sqrt{\kappa/\Gamma}\,|\bm{q}|$.

In contrast, amplitude (Higgs) modes~\cite{Higgs-o,Higgs_Anderson,Higgs_condensed_Pekker,Higgs_SC_Naoto} -- associated with fluctuations in the magnitude of the order parameter -- remain gapped.
Furthermore, in multi-Q phases, such as the triple-Q 120$^\circ$ phase in the $O(2)$ model, the coupling between different $\bm{Q}_l$ channels and the underlying $D_3$ lattice symmetry leads to hybridization of modes (Leggett modes~\cite{Leggett-o,Leggett_optical_Naoto,Goldstone_Leggett_Takashi}) and richer excitation spectra. 

In total, the $O(N)$ model exhibits $N$, $2N$, and $3N$ collective modes for single-Q, double-Q, and triple-Q phases, respectively, as summarized in Table~\ref{tab:phases}.
These gapped modes can be classified according to the residual symmetry of the corresponding ordered phase. An amplitude mode, denoted by the irreducible representation $A$ or $A_1$ (refer to Table~\ref{tab:phases}), is consistently observed to present a universal excitation gap of $\sqrt{-2\alpha/\Gamma}$.
These features can be probed by techniques such as inelastic neutron scattering~\cite{Inelastic_neutron_ScatteringBook,klotz_inelastic_2000} and Raman spectroscopy~\cite{Raman_Spectroscopy}.
It should be emphasized that considering the spin-lattice coupling will further reduce the residual symmetries and lift the degeneracy of the collective modes.

\begin{table}[tb]
	\setlength{\tabcolsep}{1.3ex}
	\renewcommand{\arraystretch}{1.5}
	\caption{Summary of phases, residual symmetries (Res. Symm.) of ordered phases, and collective excitations for $O(2)$ and $O(3)$ models. The collective modes are classified according to the irreducible representations of the residual symmetry group's discrete component~\cite{Koster-Point-Groups}. For the $O(3)$ model, the residual internal $O(2)$ symmetry introduces an extra two-fold degeneracy. }
	\label{tab:phases}
	\begin{tabular}{c|c|ccc}
		\hline\hline
		Model & Phase & Res. Symm. & Degeneracy & Modes\\
		\hline
		& $\mathrm{I}_{O(2)}$ & $C_2\times{}\mathbb{Z}_2$ & 1 (gapped) & $A$ \\
		\cline{4-5}
		& & & 1 (gapless) & $B$ \\
		\cline{2-5}
		& $\mathrm{II}^B_{O(2)}$ & $C_{4}$ & 1 (gapped) & $A$\\
		$O(2)$ & Orthogonal &   & 1 (gapped) & $B$ \\
		& & & 1 (gapped) & $E$\\
		\cline{4-5}
		& & & 1 (gapless) & $E$ \\
		\cline{2-5}
		& $\mathrm{III}^A_{O(2)}$ &  $D_3\times{}\mathbb{Z}_2$  & 1 (gapped) & $A_1$ \\
		& Collinear & & 2 (gapped) & $E$ \\
		& & & 2 (gapped) & $E$ \\
		\cline{4-5}
		& & & 1 (gapless) & $A_1$ \\
		\cline{2-5}
		& $\mathrm{III}^C_{O(2)}$ & $D_3$ & 1 (gapped) & $A_1$ \\
		& 120$^\circ$ & & 2 (gapped) & $E$ \\		
		& & & 2 (gapped) & $E$ \\
		\cline{4-5}
		& & & 1 (gapless) & $A_1$ \\
		\hline
		& $\mathrm{I}_{O(3)}$ & $C_{2}\times{}O(2)$  & 1 (gapped) & $A$ \\
		\cline{4-5}
		& & & 2 (gapless) & $A$ \\
		\cline{2-5}
		$O(3)$ & $\mathrm{III}^A_{O(3)}$ & $D_3\times{}O(2)$  & 1 (gapped) & $A_1$ \\
		& Collinear & & 2 (gapped) & $E$ \\
		& & & 4 (gapped) & $E$ \\
		\cline{4-5}
		& & & 2 (gapless) & $A_1$ \\
		\cline{2-5}
		& $\mathrm{III}^B_{O(3)}$ & $O$ & 1 (gapped) & $A_1$\\
		& Orthogonal & & 2 (gapped) & $E$ \\
		& & & 3 (gapped) & $T_2$ \\
		\cline{4-5}
		& & & 3 (gapless) & $T_1$ \\
		\hline\hline
	\end{tabular}
\end{table}

\section{Summary and Discussion}\label{sec:summary-discussion}
We have developed a comprehensive Ginzburg-Landau theoretical framework for triple-Q magnetic orders on hexagonal lattices, focusing on $O(N)$ models with $N=2$ and $N=3$. The advantage of this phenomenological theory lies in its ability to capture a broad range of behaviors through symmetry constraints alone, facilitating comparisons across different systems. Our key findings include: (i) Establishment of complete phase diagrams for both $N=2$ and $N=3$ systems, characterized by competition between interaction parameters $\beta_1$, $\beta_2$, and $\beta_3$; (ii) Identification of four distinct phases for $N=2$ [$\mathrm{I}_{O(2)}$, $\mathrm{II}^B_{O(2)}$, $\mathrm{III}^A_{O(2)}$, $\mathrm{III}^C_{O(2)}$] and three stable phases for $N=3$ [$\mathrm{I}_{O(3)}$, $\mathrm{III}^A_{O(3)}$, $\mathrm{III}^B_{O(3)}$]; and (iii) Systematic analysis of collective excitations revealing fundamental connections between symmetry breaking patterns and collective modes.

These findings have significant physical implications: (i) They provide a theoretical framework for understanding complex magnetic structures observed in hexagonal lattice compounds, encompassing triangular, honeycomb, and kagome lattices;
(ii) They offer specific predictions for experimental probes of magnetic excitations; and (iii) They establish a comprehensive classification system for symmetry-breaking patterns in these magnetic systems.

Future research directions include: 

(i) Incorporation of external fields and additional anisotropies~\cite{Messio2011,Janssen2016}. In particular, in a magnetic overlayer the point‐group symmetry is reduced from $D_3$ to $C_{3v}$ and the inversion is broken, so that Dzyaloshinskii-Moriya interactions (DMI) become symmetry-allowed. In our continuum language the leading effect of DMI is to generate Lifshitz invariants-terms linear in spatial gradients of the order parameters-of the form
\begin{equation}
\mathcal{F}_{DM} \sim\sum_{lm}D_{lm} \bm{\Delta}_{\bm{Q}_l}\cdot(\hat{z}\times\nabla)\cdot\bm{\Delta}_{\bm{Q}_m},
\end{equation}
where $\hat{z}$ is the surface normal and $D_{lm}$ are chiral coupling constants determined by the microscopic spin-orbit and exchange processes. Physically, these terms (1) favor chiral (cycloidal or helical) arrangements at each M-point rather than purely collinear order; (2) lift the degeneracy between modes at $+\bm{q}$ and $-\bm{q}$, leading to non-reciprocal magnon dispersions $\omega(\bm{q})\neq{}\omega(-\bm{q})$;
(3) Introduce anisotropy in the spin-wave velocities and can open small gaps at otherwise gapless (Goldstone) points.

(ii) Extension to other internal symmetries, such as $U(N)$ models featuring complex order parameters, is notably relevant. In particular, $U(1)$ models are intimately linked to the recently observed triple-Q pair density wave in the Kagome superconductor family AV$_3$Sb$_5$ (A=K, Rb or Cs)~\cite{chen2021,Jin2022,Zhou2022,yao2024}. Since $O(2)\cong{}U(1)$, our $N=2$ model can be expanded to encompass systems characterized by three complex scalar order parameters, $\Delta_{\bm{Q}_l}$ for $l=1,\,2,\,3$, effectively forming a $U(1)$ model. Further explanations are provided in Appendix~\ref{app:U1}. 

(iii) Phase transition dynamics~\cite{Hohenberg1977}, topological defects and excitations~\cite{Mermin1979} based on the multiple-Q formulation.

(iv) Application to specific material systems with experimental validation.

Finally, we would like to remark that while the coefficients of our phenomenological theory are treated as general parameters, they can be mapped to microscopic energy scales in specific material systems or spin models, offering a bridge between universal symmetry-based predictions and material-specific calculations.

\acknowledgments
We would like to thank Yuan Li and Hong Yao for their helpful discussions. This work is supported in part by the National Key Research and Development Program of China (Grant No. 2022YFA1403403), the National Natural Science Foundation of China (Grants No. 12274441 and No. 12034004). J.-T.J. is supported by a fellowship and a CRF award from the Research Grants Council of the Hong Kong Special Administrative Region, China (Projects No. HKUST SRFS2324-6S01 and No. C703722GF). J.-T.J. also acknowledge the support from the New Cornerstone Science Foundation.

\appendix

\section{Effective Lagrangian}\label{app:effective Lagrangian}

In this section, we  give the derivation of the symmetry allowed effective Lagrangian $\mathcal{L}_{O(N)}$ as well as the corresponding Euler-Lagrange equation. 

In general, the effective Lagrangian $\mathcal{L}_{O(N)}$, including both temporal and spatial derivatives, is given by:
\begin{equation}
\begin{aligned}
\mathcal{L}_{O(N)} &= \sum_{pqlm} \Gamma^{1}_{pqlm} \left( \Delta^p_{\bm{Q}_l} \partial_t \Delta^q_{\bm{Q}_m} - \Delta^q_{\bm{Q}_m} \partial_t \Delta^p_{\bm{Q}_l} \right)\\
&+ \sum_{pqlm} \Gamma^2_{pqlm} \left( \partial_t \Delta^p_{\bm{Q}_l} \right) \left( \partial_t \Delta^q_{\bm{Q}_m} \right) \\
&- \sum_{\mu\nu pqlm} \kappa_{\mu\nu pqlm} \left( \partial_\mu \Delta^p_{\bm{Q}_l} \right) \left( \partial_\nu \Delta^q_{\bm{Q}_m} \right) - \mathcal{F}_{O(N)},
\end{aligned}
\end{equation}    
where $\Gamma^{1(2)}_{pqlm}$ and $\kappa_{\mu\nu pqlm}$ are real coefficients with indices $p,q=1,2,\dots,N$ and $l,m=1,2,3$, while $\mu,\nu = x,y$. This Lagrangian $\mathcal{L}_{O(N)}$ needs to remain invariant under the symmetry operations of the system, imposing restrictions on these constants. This symmetry includes: (i) translational symmetry, (ii) internal $O(N)$ symmetry, and (iii) $D_3$ rotational symmetry.

The translational symmetry requires that the total momentum of each term in $\mathcal{L}_{O(N)}$ vanishes (up to a reciprocal lattice vector), resulting in $\Gamma^{1(2)}_{pqlm}=\Gamma^{1(2)}_{pql}\delta_{lm}$ and $\kappa_{\mu\nu{}pqlm}=\kappa_{\mu\nu{}pql}\delta_{lm}$. In addition, the $O(N)$ symmetry allows the exchange of two components in $\bm{\Delta}_{\bm{Q}_l}$, i.e., $\Delta^p_{\bm{Q}_l} \leftrightarrow \Delta^q_{\bm{Q}_l}$. Thus, $\Gamma^{1(2)}_{pql}$ and $\kappa_{\mu\nu{}pql}$ must be symmetric under the exchange of indices $p$ and $q$. However, it is easy to see that $\Delta^p_{\bm{Q}_l}\partial_t\Delta^q_{\bm{Q}_l}-\Delta^q_{\bm{Q}_l}\partial_t\Delta^p_{\bm{Q}_l}$ changes sign under the transformation $\Delta^p_{\bm{Q}_l} \leftrightarrow \Delta^q_{\bm{Q}_l}$. Thus $\Gamma^{1}_{pql}$ must be zero. Furthermore, the $O(N)$ symmetry also enforces that the $\Gamma^2$ and $\kappa$ related terms to $O(N)$-invariant inner products, i.e., $\sum_{l}\Gamma^2_{l}(\partial_t\bm{\Delta}_{\bm{Q}_l})\cdot(\partial_t\bm{\Delta}_{\bm{Q}_l})$ and$
-\sum_{\mu\nu{}l}\kappa_{\mu\nu{}l}(\partial_\mu\bm{\Delta}_{\bm{Q}_l})\cdot(\partial_\nu\bm{\Delta}_{\bm{Q}_l})$.

Then we need to consider the $D_3$ symmetry. Since the $D_3$ operations do not change the temporal derivative $\partial_t$, the constraint for $\Gamma^{2}$ related terms is simple: $\Gamma^{2}_{1}=\Gamma^{2}_{2}=\Gamma^{2}_{3}\equiv{}\Gamma$. As for the $\kappa$ related terms, there are the following $9$ inequivalent terms in total,
\begin{widetext}
 \begin{align*}
t_1 &= (\partial_x\bm{\Delta}_{\bm{Q}_1})\cdot(\partial_x\bm{\Delta}_{\bm{Q}_1}),\quad
t_2 = (\partial_x\bm{\Delta}_{\bm{Q}_2})\cdot(\partial_x\bm{\Delta}_{\bm{Q}_2}),\quad
t_3 = (\partial_x\bm{\Delta}_{\bm{Q}_3})\cdot(\partial_x\bm{\Delta}_{\bm{Q}_3}),\\[1mm]
t_4 &= (\partial_x\bm{\Delta}_{\bm{Q}_1})\cdot(\partial_y\bm{\Delta}_{\bm{Q}_1}),\quad
t_5 = (\partial_x\bm{\Delta}_{\bm{Q}_2})\cdot(\partial_y\bm{\Delta}_{\bm{Q}_2}),\quad
t_6 = (\partial_x\bm{\Delta}_{\bm{Q}_3})\cdot(\partial_y\bm{\Delta}_{\bm{Q}_3}),\\[1mm]
t_7 &= (\partial_y\bm{\Delta}_{\bm{Q}_1})\cdot(\partial_y\bm{\Delta}_{\bm{Q}_1}),\quad
t_8 = (\partial_y\bm{\Delta}_{\bm{Q}_2})\cdot(\partial_y\bm{\Delta}_{\bm{Q}_2}),\quad
t_9 = (\partial_y\bm{\Delta}_{\bm{Q}_3})\cdot(\partial_y\bm{\Delta}_{\bm{Q}_3}).
\end{align*}   
\end{widetext}
The spatial derivative terms can be written as $\mathcal{L}_{\kappa}=-\sum_{i=1}^9{}\kappa_i{}t_i$ where $\kappa_i$ is the corresponding strength of $t_i$.

Now we apply the $D_3$ operators to $\mathcal{L}_\kappa$. If $\mathcal{L}_\kappa$ is $D_3$ symmetric, all the $D_3$ operators keep it unchanged. This analysis will give rise to constraints on the $9$ coefficients $\kappa_i$. We consider the two generators, $\mathcal{A}$ and $\mathcal{D}$, of $D_3$ group. They act on momenta and derivatives as:
\begin{align*}
\mathcal{A}(\bm{Q}_1,\,\bm{Q}_2,\,\bm{Q}_3)&=(\bm{Q}_1,\,\bm{Q}_3,\,\bm{Q}_2),\\
\mathcal{D}(\bm{Q}_1,\,\bm{Q}_2,\,\bm{Q}_3)&=(\bm{Q}_2,\,\bm{Q}_3,\,\bm{Q}_1),\\
\mathcal{A}(\partial_x,\,\partial_y)&=(-\partial_x,\,\partial_y),\\
\mathcal{D}(\partial_x,\,\partial_y)&=(\frac{1}{2}\partial_x+\frac{\sqrt{3}}{2}\partial_y,\,-\frac{\sqrt{3}}{2}\partial_x+\frac{1}{2}\partial_y).
\end{align*}
It is easy to check that 
\begin{align*}
\mathcal{A}t_1&=t_1,\,\mathcal{A}t_2=t_3,\,\mathcal{A}t_3=t_2,\\
\mathcal{A}t_4&=-t_4,\,\mathcal{A}t_5=-t_6,\,\mathcal{A}t_6=-t_5,\\
\mathcal{A}t_7&=t_7,\,\mathcal{A}t_8=t_9,\,\mathcal{A}t_9=t_8,
\end{align*}
Thus, the condition $\mathcal{A}\mathcal{L}_\kappa=\mathcal{L}_\kappa$ requires that $\kappa_2=\kappa_3,\,\kappa_5=-\kappa_6,\,\kappa_8=\kappa_9$ and $\kappa_4=0$. By imposing these constraints, $\mathcal{L}_\kappa=-\kappa_1{}t_1-\kappa_2(t_2+t_3)-\kappa_5(t_5-t_6)-\kappa_7{}t_7-\kappa_8(t_8+t_9)$. Then we can further apply the operator $\mathcal{D}$ to each single term as follows,
\begin{align*}
\mathcal{D}t_1&=\frac{1}{4}t_2+\frac{3}{4}t_8+\frac{\sqrt{3}}{4}t_5,\\
\mathcal{D}(t_2+t_3)&=\frac{1}{4}(t_1+t_3)+\frac{3}{4}(t_7+t_9)+\frac{\sqrt{3}}{4}(t_4+t_5),\\
\mathcal{D}(t_5-t_6)&=\frac{\sqrt{3}}{4}(t_1-t_3-t_7+t_9)+\frac{1}{2}(t_4-t_6),\\
\mathcal{D}t_7&=\frac{3}{4}t_2+\frac{1}{4}t_8-\frac{\sqrt{3}}{4}t_5,\\
\mathcal{D}(t_8+t_9)&=\frac{3}{4}(t_1+t_3)+\frac{1}{4}(t_7+t_9)-\frac{\sqrt{3}}{4}(t_4+t_5).
\end{align*}
Now the condition $\mathcal{D}\mathcal{L}_\kappa=\mathcal{L}_\kappa$ can be expressed as 
\begin{align*}
\kappa_1&=\frac{1}{4}(\kappa_2+\sqrt{3}\kappa_5+3\kappa_8),\\
\kappa_2&=\frac{1}{4}(\kappa_1+3\kappa_7),\\
\kappa_3=\kappa_2&=\frac{1}{4}(\kappa_2-\sqrt{3}\kappa_5+3\kappa_8),\\
\kappa_4=0&=\frac{\sqrt{3}}{4}(\kappa_2-\kappa_8)+\frac{1}{2}\kappa_5,\\
\kappa_5&=\frac{\sqrt{3}}{4}(\kappa_1+\kappa_2-\kappa_7-\kappa_8),\\
\kappa_6=-\kappa_5&=-\frac{1}{2}\kappa_5,\\
\kappa_7&=\frac{1}{4}(3\kappa_2-\sqrt{3}\kappa_5+\kappa_8),\\
\kappa_8&=\frac{1}{4}(3\kappa_1+\kappa_7),\\
\kappa_9=\kappa_8&=\frac{1}{4}(3\kappa_2+\sqrt{3}\kappa_5+\kappa_8).
\end{align*}
The above equations give rise to $\kappa_1=\kappa_2=\kappa_7=\kappa_8\equiv{}\kappa$ and $\kappa_5=0$. Thus, $\mathcal{L}_\kappa$ simplifies to
\begin{align*}
\mathcal{L}_\kappa&=-\kappa(t_1+t_2+t_3+t_7+t_8+t_9)\\
&=-\kappa\sum_l\left[(\partial_x\bm{\Delta}_{\bm{Q}_l})\cdot(\partial_x\bm{\Delta}_{\bm{Q}_l})+(\partial_y\bm{\Delta}_{\bm{Q}_l})\cdot(\partial_y\bm{\Delta}_{\bm{Q}_l})\right].
\end{align*}    
Finally, the effective Lagrangian becomes
\begin{widetext}
\begin{equation}
\begin{aligned}
\mathcal{L}_{O(N)}&=
\Gamma\sum_{l}(\partial_t\bm{\Delta}_{\bm{Q}_l})\cdot(\partial_t\bm{\Delta}_{\bm{Q}_l})-\kappa\sum_{l}\left[(\partial_x\bm{\Delta}_{\bm{Q}_l})\cdot(\partial_x\bm{\Delta}_{\bm{Q}_l})+(\partial_y\bm{\Delta}_{\bm{Q}_l})\cdot(\partial_y\bm{\Delta}_{\bm{Q}_l})\right]
-\mathcal{F}_{O(N)},\\
&\approx
\Gamma\sum_{l}(\partial_t\bm{\eta}_{\bm{Q}_l})\cdot(\partial_t\bm{\eta}_{\bm{Q}_l}) - \kappa\sum_{l}\left[(\partial_x\bm{\eta}_{\bm{Q}_l})\cdot(\partial_x\bm{\eta}_{\bm{Q}_l})+(\partial_y\bm{\eta}_{\bm{Q}_l})\cdot(\partial_y\bm{\eta}_{\bm{Q}_l})\right]-\mathcal{F}_{O(N)},
\end{aligned}
\end{equation}     
\end{widetext}   
where $\Gamma$ and $\kappa$ are positive constants that represent the kinetic and spatial coupling parameters, respectively. This is exact Eq.~\eqref{eq:effective-lagrangian}.

The dynamics of the fluctuations are governed by the Euler-Lagrange equation, which is given as follows:
\begin{equation}
\begin{aligned}
&\partial_t\left(\frac{\partial\mathcal{L}_{O(N)}}{\partial(\partial_{t}\bm{\eta}_{\bm{Q}_l})}\right)+\partial_x\left(\frac{\partial\mathcal{L}_{O(N)}}{\partial(\partial_{x}\bm{\eta}_{\bm{Q}_l})}\right)+\partial_y\left(\frac{\partial\mathcal{L}_{O(N)}}{\partial(\partial_{y}\bm{\eta}_{\bm{Q}_l})}\right)\\
=&\frac{\partial\mathcal{L}_{O(N)}}{\partial\bm{\eta}_{\bm{Q}_l}}=-\frac{\partial[\delta\mathcal{F}_{O(N)}]}{{\partial\bm{\eta}_{\bm{Q}_l}}}.      
\end{aligned}
\end{equation}
It gives rise to the equation of motion Eq.~\eqref{eq:collective-mode-equation}.

\section{$U(1)$ model}\label{app:U1}

The results for $N=2$ can help us study certain systems described by three complex scalar order parameters $\Delta_{\bm{Q}_l}$($l=1,\,2,\,3$) on hexagonal lattice. We give a brief discussion on this point in this section.

In additional to  D$_3$ rotational symmetry and lattice translational symmetry, we consider a U($1$) symmetric theory. Here the U($1$) symmetry means that for any transformation $\Delta_{\bm{Q}_l} \mapsto \Delta_{\bm{Q}_l}e^{i\theta}$, the free energy $\mathcal{F}_{U(1)}$ of the system unchanged, where $\theta$ is a real number. For such a system, $\mathcal{F}_{U(1)}$ is of the following form,

\begin{equation}\label{eq-sm:U(1)-F-0}
\begin{aligned}
\mathcal{F}_{U(1)} &=  \alpha^{U(1)} \sum_{l=1}^{3} |{\Delta}_{{\bm{Q}}_l}|^2
+ \beta_1^{U(1)} \sum_{l=1}^{3} |\Delta_{{\bm{Q}}_l}|^4\\
&+\beta_2^{U(1)} \sum_{l<m}|\Delta_{{\bm{Q}}_l}|^2|\Delta_{{\bm{Q}}_m}|^2\\
&+ \beta_3^{U(1)} \sum_{l<m}\left(\Delta_{{\bm{Q}}_l} \Delta^{*}_{{\bm{Q}}_m}\right)^2+c.c..
\end{aligned}
\end{equation}

We can use two values $|\Delta_{\bm{Q}_l}|$ and $\theta_l$ to define each $\Delta_{\bm{Q}_l}$, i.e.,  $\Delta_{\bm{Q}_l}=|\Delta_{\bm{Q}_l}|e^{i\theta_l}$. Then the $\beta^{U(1)}_3$ related terms in Eq.~\eqref{eq-sm:U(1)-F-0} can be expressed as 

\begin{equation}\label{eq-sm:U(1)-F-beta-3}
\begin{aligned}
&\beta_3^{U(1)} \sum_{l<m}\left(\Delta_{{\bm{Q}}_l} \Delta^{*}_{{\bm{Q}}_m}\right)^2 + c.c.\\
= &2\beta^{U(1)}_{3}\sum_{l<m}|\Delta_{{\bm{Q}}_l}|^2|\Delta_{{\bm{Q}}_m}|^2\cos(2\theta_l-2\theta_m).
\end{aligned}
\end{equation} 
By comparing $\mathcal{F}_{U(1)}$ with $\mathcal{F}_{O(2)}$, we can obtain the properties of $\mathcal{F}_{U(1)}$ through the following mapping,
\begin{equation*}
\begin{aligned}
|\bm{\Delta}_{{\bm{Q}}_l}|\mapsto|\Delta_{{\bm{Q}}_l}|,& \quad \phi_{l}\mapsto \theta_l,\\
\alpha \mapsto \alpha^{U(1)},&\quad
\beta_1 \mapsto \beta_1^{U(1)},\\
\beta_2 +\frac{\beta_3}{2}\mapsto \beta_2^{U(1)},&\quad 
\frac{\beta_3}{4}\mapsto \beta_3^{U(1)}.
\end{aligned}
\end{equation*} 

\begin{figure}[tb]
\includegraphics[width=0.8\linewidth]{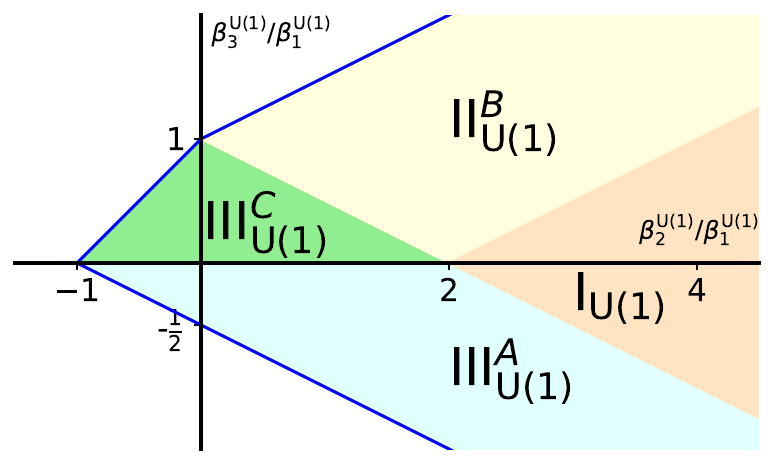}
\caption{The phase diagram and the stable region for $\mathcal{F}_{U(1)}$.}\label{fig:phase-diagram-U(1)}
\end{figure}

From this mapping, we can obtain the phase diagram of $\mathcal{F}_{U(1)}$. The result is shown in  Fig.~\ref{fig:phase-diagram-U(1)}. The phases $\mathrm{I}_{O(2)}$, $\mathrm{II}^{B}_{O(2)}$ and $\mathrm{III}^{A(C)}_{O(2)}$ are mapped to $\mathrm{I}_{U(1)}$, $\mathrm{II}^{B}_{U(1)}$ and $\mathrm{III}^{A(C)}_{U(1)}$, respectively. The same mapping will also give rise to the stable conditions: ${\beta}^{U(1)}_2-2{\beta}^{U(1)}_3+2\beta^{U(1)}_1>0$, ${\beta}^{U(1)}_2+2{\beta}^{U(1)}_3+\beta^{U(1)}_1>0$ and ${\beta}^{U(1)}_2-{\beta}^{U(1)}_3+\beta^{U(1)}_1>0$, along with the conditions $\alpha^{U(1)}<0$ and $\beta^{U(1)}_1>0$.

\bibliography{triple-Q-GL}

\end{document}